  \documentstyle[12pt]{article}
  
\catcode`@=11
\def\secteqno{\@addtoreset{equation}{section}%
\def\theequation{\thesection.\arabic{equation}}}
\catcode`@=12
\secteqno
\newcommand{\be}{\begin{equation}}
\newcommand{\ee}{\end{equation}}
\newcommand{\bea}{\begin{eqnarray}}
\newcommand{\eea}{\end{eqnarray}}
\newcommand{\bref}[1]{(\ref{#1})}
\newcommand{\nn}{\nonumber}

\newcommand{\slbL}{/ {\hskip-0.27cm{\bf L}}}

\newcommand{\slbmpi}
{/ {\hskip-0.27cm{\mbox{\boldmath $\pi$}}}}
\newcommand{\slbmx}
{/ {\hskip-0.27cm{\mbox{\boldmath $x$}}}}
\newcommand{\lcpm}{\varpi}
\newcommand{\uu}{u}
\newcommand{\vv}{v}
\newcommand{\bmx}{\mbox{\boldmath $x$}}
\newcommand{\bmp}{\mbox{\boldmath $p$}}
\newcommand{\bma}{\mbox{\boldmath $\alpha$}}
\newcommand{\bmpi}{\mbox{\boldmath $\pi$}}
\newcommand{\hepth}{hep-th}

\begin{document}
\thispagestyle{empty}
\vfill
\hfill July 17, 2002\par
\hfill KEK-TH-835\null\par
\hfill TOHO-CP-0272\null\par
\vskip 20mm
\begin{center}
{\Large\bf Gauge invariant action for superstring}\par
\vskip 6mm
{\Large\bf in Ramond-Ramond plane-wave background}\par
\vskip 6mm
\medskip

\vskip 10mm
{\large Machiko\ Hatsuda,~~Kiyoshi~Kamimura$^\dagger$~and~~Makoto\ Sakaguchi }\par
\medskip
{\it 
Theory Division,\ High Energy Accelerator Research Organization (KEK),\\
 Tsukuba,\ Ibaraki,\ 305-0801, Japan }\\
{\it 
$~^\dagger$ 
 Department of Physics, Toho University, Funabashi, 274-8510, Japan}\\
\medskip\medskip
{\small\sf E-mails:\ mhatsuda@post.kek.jp, kamimura@ph.sci.toho-u.ac.jp,
Makoto.Sakaguchi@kek.jp} 
\medskip
\end{center}
\vskip 10mm
\begin{abstract}
We present a gauge invariant action for a superstring in 
the plane wave background with Ramond-Ramond (RR) five-form flux.
The Wess-Zumino term is given explicitly
in a bilinear form of the 
left invariant currents by introducing a
fermionic center to define the nondegenerate group metric.  
The reparametrization invariance generators, 
whose combinations are conformal generators, 
and fermionic
constraints, half of which generate $\kappa$-symmetry, 
are obtained. 
Equations of motion are obtained in 
conformal invariant and background covariant manners.
\end{abstract} 
\par\par
\newpage
\setcounter{page}{1}
\parskip=7pt
\section{ Introduction}\par
\indent

Recently the plane wave solution with the Ramond-Ramond (RR) 5-form flux
was found as a maximally supersymmetric type IIB supergravity solution
\cite{joseIIB} 
in addition to the Minkowski flat and the AdS$_5$$\times$S$^5$ 
spaces,
based on studies of the plane wave solutions with the 4-form flux
in the 11-dimensional supergravity \cite{KG1}.
The Penrose's limiting procedure 
\cite{Penrose} was applied to the AdS spaces to 
obtain these plane wave solutions with fluxes (pp-wave) 
\cite{joseP,HKSpp,HKSppm}.
It is recognized as an approximation of 
AdS spaces and leads to interesting approaches to the 
AdS/CFT correspondence \cite{Malda}.

The superstring action in the RR plane wave background 
was presented \cite{Me1} and 
it was also shown that the action in the light-cone gauge
becomes simply an action for 8 bosons and 8 fermions 
which are free  and  massive in 2-dimensions.
Brane actions in the RR pp-wave background have been widely
studied \cite{Dab} mostly in the light-cone gauge.
In the light-cone gauge the conformal symmetry 
is broken by a 2-dimensional mass term at the 
gauge fixed level though it should be recovered 
in whole string theory even in the RR pp-wave background \cite{Takay}.
The light-cone Hamiltonian does not commute with 
other global space-time charges thus 
states in a supergravity multiplet have different 
light-cone energy values \cite{MeTsy2}
and the light-cone energy is not minimized for BPS states.
From a point of view of the symmetry, the 
light-cone approach is not suitable to understand systems. 
Manifest conformal invariant approaches have been providing us 
elegant formulations of string theories and practical computation
methods.
However the conformal invariant treatment has not been explored
except in an alternative hybrid approach \cite{Berkp}.
In this paper we will study the Green-Schwarz type superstring action 
in the RR pp-wave background in a covariant and manifest conformal 
invariant way.

The RR backgrounds are usually described by the
Green-Schwarz type actions which contain Wess-Zumino (WZ) terms.
In reference \cite{Me1} the WZ term of the superstring in pp-wave 
background is given in 
a one parameter integral of a closed three form. 
The formal integral representation of the WZ term is often useful in
discussing the invariance of the action but is not always convenient for
the practical analysis.
The integral is hardly performed especially for curved background cases
except in the light-cone gauge,
so local symmetry constraints and covariant equations of motion 
which will be needed in the covariant
string field theories are hardly discussed.
On the other hand,
it was shown that the WZ term
for the covariant superstring in AdS spaces can be constructed 
in a bilinear form of the left-invariant (LI) currents 
\cite{Berk,HKS,RWS,AdSWZ},
and local symmetry constraints \cite{HKAdS} and global charges \cite{AdSWZ}
were obtained explicitly.
For the super-AdS group
the WZ term which is bilinear in the LI currents 
can be constructed using
a nondegenerate group metric depending on
the scale parameter.
In this paper we obtain 
the superstring action in the super-pp wave
background from the one in the super-AdS background \cite{AdSWZ} 
using the  Penrose limit \cite{HKSpp}.
The limiting procedure must be taken carefully,
since the bilinear form WZ term contains
a divergent term when the Penrose limit parameter, $\Omega$ , 
is brought to zero.
In a flat limit where the AdS radius goes to infinity,
the bilinear form WZ term also contains a divergent term 
which is a total derivative term causing no difficulty.
On the other hand the divergent term in the Penrose limit, 
which is a bilinear product of leading terms of 
$\Omega$ power series of LI currents,
is not a total derivative. 
The correct form of the billinear WZ term is obtained
by subtracting the divergent term 
and preserving a next to leading term 
of LI currents that gives a finite contribution.
The next to leading term corresponds to the fermionic center 
introduced to make the group metric nondegenerate \cite{green}.

The organization of this paper is as followings.
The Penrose limit of the bosonic sting is explained 
as a simpler case in the section 2.
In the section 3, 
the correct limiting procedure of the bilinear form WZ term
is explained. The resulting WZ term is shown to produce the closed 
three form $H_{[3]}$.
The $\kappa$-invariance  of the action is  also confirmed. 
Then the gauge invariant action for the RR pp-wave background 
is presented. 
It is also shown that 
this action reduces to the light-cone action obtained by Metsaev \cite{Me1}
and the Green-Schwarz action \cite{GS} in the flat limit.
In the section 4, 
we examine a particle, 
a superparticle, a bosonic string and a superstring cases.
For each systems generators of reparametrizations
and fermionic constraints are calculated.
We obtain Hamiltonians and equations of motion 
in the conformal gauge.

\section{ Penrose limit of bosonic string action}\par
\indent

The superstring in the AdS$_5\times$S$^5$ space is
given as a
sigma model action on 
$\frac{SU(2,2\mid 4)}{SO(4,1)\times SO(5)}$ \cite{MeTsy},
and we begin with a bosonic  part of
it. 
We use the same notation of the superalgebra as  \cite{MeTsy,AdSWZ}.
Bosonic part of the LI Cartan 1-forms are given by 
$G_{\rm AdS}(y)^{-1}dG_{\rm AdS}(y)=e^a_{\rm AdS} P_a+e^{a'}_{\rm AdS}P_{a'}
+\frac{1}{2}\omega^{ab}_{\rm AdS}M_{ab}+
\frac{1}{2}\omega^{a'b'}_{\rm AdS}M_{a'b'}$  
where we use the subscript ``AdS" to indicate 
objects on the AdS$_5\times$S$^5$ space.
Parametrizing an element of the coset as 
$G_{\rm AdS}(y)={\rm exp}(y^{\hat{a}}P_{\hat{a}})$, $\hat{a}=(a,a')$, 
it follows the Cartan 1-forms as 
\bea
e^a_{\rm AdS}=dy^a+\left(\displaystyle\frac{\sinh y}{y}-1\right)dy^b{\bf \Upsilon}_b^{~~a}
&,&
e^{a'}_{\rm AdS}=dy^{a'}+\left(\displaystyle\frac{\sin y'}{y'}-1\right)dy^{b'}{\bf \Upsilon}_{b'}^{~~a'} \label{AdSe}
\eea
with
\bea
y&=&\sqrt{y^2}=\sqrt{y^ay_a},~~
y'~=~\sqrt{{y'}^2}=\sqrt{{y}^{a'}y_{a'}},\nn\\
{\bf \Upsilon}_{a}^b&=&\delta_a^{~b}-\displaystyle\frac{y_ay^b}{y^2}~~,
{\bf \Upsilon}_{a'}^{b'}~=~\delta_{a'}^{~b'}-\displaystyle\frac{y_{a'}y^{b'}}
{{y'}^2}~~\label{projectionsb}~~~.
\eea
A scale parameter $R$, 
which is the radius of AdS$_5$ and S$^5$, is introduced 
in the normalization of
$F_5=(1/2R)({\rm dvol} ({\rm AdS}_5)+{\rm dvol} ({\rm S}^5))$,
by rescaling $P_{\hat{a}} \to RP_{\hat{a}}$.   
The bosonic part of the sigma model action is written as
\bea
{\cal L}_{0,\rm AdS}&=&e^{\hat{a}}_{\rm AdS}e_{{\rm AdS},\hat{a}}
=
(dy)^2+(\sinh \frac{y}{R})^2d\Omega_4^2+(dy')^2
+(\sin \frac{y'}{R})^2d'\Omega^2_{4'}
~~
\eea
where $d\Omega_4^2$ and $d\Omega^2_{4'}$ are 4-sphere metrics.

The Penrose limit is obtained as $\Omega \to 0$ after rescaling 
\bea
y^+\to \Omega^2 y^+,~~y^-\to y^-,~~y^{\hat{i}}\to \Omega y^{\hat{i}}
\label{Penvar}
\eea
where $y^\pm=(y^9 \pm y^0)/\sqrt{2}$,
corresponding to $
P_+\to \Omega^{-2} P_+,~~P_-\to P_-,~~P_{\hat{i}}\to \Omega^{-1} P_{\hat{i}}
$. 
Rewriting \bref{AdSe} in terms of \bref{Penvar} and 
taking $\Omega$ infinitesimally small, 
 \bref{AdSe} turns out to be a power series of $\Omega$.
Cartan 1-forms should be also rescaled as
$e^+\to \Omega^2 e^+,~~e^-\to e^-,~~e^{\hat{i}}\to \Omega e^{\hat{i}}$ 
for consistency.
Taking $\Omega \to 0$ limit, leading terms of the expansion in $\Omega$
are identified to the ones in the plane wave background
\bea
\left\{\begin{array}{lcl}
\Omega^2 e^+_{\rm AdS}&=&\Omega^2\left[dy^+
+\left(\displaystyle\frac{\sin \frac{y^-}{\sqrt{2}R}}{\frac{y^-}{\sqrt{2}R}}-1\right)
\displaystyle\frac{y^{\hat{i}}}{y^-}\left(y^{\hat{i}}\displaystyle\frac{dy^-}{y^-}-dy^{\hat{i}}\right)
\right]+o(\Omega^4)\equiv \Omega^2 e_{\rm pp}^++o(\Omega^4)\\
e^{-}_{\rm AdS}&=&dy^-+o(\Omega^4)\equiv e_{\rm pp}^-+o(\Omega^4)
\\
\Omega e^{\hat{i}}_{\rm AdS}&=&\Omega\left[
dy^{\hat{i}}-\left(
\displaystyle\frac{\sin \frac{y^-}{\sqrt{2}R}}{\frac{y^-}{\sqrt{2}R}}-1\right)
\left(y^{\hat{i}}\displaystyle\frac{dy^-}{y^-}-dy^{\hat{i}}\right)
\right]+o(\Omega^3)\equiv \Omega e_{\rm pp}^{\hat{i}}+o(\Omega^3).
~~~~~~~~~~\end{array}\right.\label{ppe}
\eea
Bosonic part of the sigma model action in the pp-wave background is written as
\bea
{\cal L}_{0,\rm pp}&=&e^{\hat{a}}_{\rm pp}e_{{\rm pp},\hat{a}}\nn\\
&=&
2dy^+ dy^-
+\left\{\left(\displaystyle\frac{\sin \frac{y^-}{\sqrt{2}R}}{\frac{y^-}
{\sqrt{2}R}}\right)^2-1\right\}
\left\{  \left(\displaystyle\frac{y^{\hat{i}}}{y^-}\right)^2
(d y^- )^2
-2\left(\displaystyle\frac{y^{\hat{i}}}{y^-}\right)
d y^{\hat{i}} d y^-
\right\}
\nn\\
&&~~~~~~~~~~~~~~~~
+\left(\displaystyle\frac{\sin \frac{y^-}{\sqrt{2}R}}
{\frac{y^-}{\sqrt{2}R}}\right)^2
d y^{\hat{i}}d y^{\hat{i}}
\nn\\
&=&2dx^+ d x^-
-\displaystyle\sum_{\hat{i}=1}^{8}
(2\mu{x^{\hat{i}}})^2
(d x^-)^2+\displaystyle\sum_{\hat{i}=1}^{8}
d x^{\hat{i}} d x^{\hat{i}},
\label{ppmetric}
\eea
where the last expression is obtained by the following field redefinition
\bea
\left\{\begin{array}{lcl}
x^{\hat{i}}&=&\displaystyle\frac{\sin 2\mu y^-}{2\mu y^-}
y^{\hat{i}}\\
x^-&=&y^-\\
x^+&=&y^+
+\displaystyle\sum_{\hat{i}=1}^{8}
\displaystyle\frac{(y^{\hat{i}})^2}{2y^-}
\left(1-
\displaystyle\frac{\sin 4\mu y^-}{4\mu y^-}
\right).
\end{array}\right.\label{yandx}
\eea
A scale parameter $\mu$ is introduced in the normalization 
$F_5=\mu dx^-\wedge( dx^1dx^2dx^3dx^4+dx^5dx^6dx^7dx^8)$ 
.
This coordinate system is obtained 
if one begins with the 
coset parameterization 
 $G_{\rm pp}(x)=e^{x^+P_+ }e^{x^-P_- }e^{x^{\hat{i}}P_{\hat{i}} }$.

\section{ Penrose limit of superstring action  }\par
\indent
\subsection{Wess-Zumino term and nondegenerate super-pp-wave algebra}\par
\indent
For an In$\ddot{\rm o}$n$\ddot{\rm u}$-Wigner group contraction \cite{IW} 
the Cartan 1-forms are expanded with respect to
a parameter $s$ which is brought to zero as 
\bea
T_A \to s^{-N_A} T_A~~\Rightarrow~~
L^A(z) \to s^{N_A}L^A(s^{N_B}z^{B})
=\displaystyle\sum_{N=N_A}^{\infty} s^N L^A_{(N)}(z)~~~,\label{exps}
\eea
as was seen in \bref{ppe} for  bosonic part.
The Maurer-Cartan equations are also expanded as
\bea
\displaystyle\sum_{N=N_A}^{\infty} s^N d L^A_{(N)}
+\frac{1}{2}f_{BC}^A
\sum_{M=N_B}^{\infty}\sum_{K=N_C}^{\infty} s^{M+K} L^B_{(M)} L^C_{(K)}=0~~~,
\label{mcss}
\eea
where $f_{BC}^A$ is the structure constant of the original Lie algebra.
Usually only leading terms of Cartan 1-forms are preserved
in the limiting procedure as was seen in the previous section.
The Maurer-Cartan equations for the leading terms
\bea
&&d L^A_{(N_A)}
+\frac{1}{2}{f}_{B(N_B)~C(N_C)}^{A(N_A)}
 L^B_{(N_B)} L^C_{(N_C)}=0~~,~\nn\\
&&~~~~ \left\{
\begin{array}{ll}
{f}_{B(N_B)~C(N_C)}^{A(N_A)}=f_{BC}^A~&,~{\rm for}~~N_A=N_B+N_C \\
{f}_{B(N_B)~C(N_C)}^{A(N_A)}=0~&,~{\rm for}~~N_A\ne N_B+N_C
\end{array}
\right.~~~\label{IWmc}
\eea
describe the resultant group structure.

The Penrose limit 
from the super-AdS group to the super-pp wave group
makes the super-pp group metric to be degenerate.
This is the similar situation as in the flat limit from the super-AdS group 
to the super-translation group where the metric is degenerate.
However if the next to leading term in 
the fermionic Cartan 1-forms \bref{exps} is maintained
in the limiting procedure,
the nondegenerate group metric of the central extended super-pp group
can be constructed
as we will show below.

The nondegenerate group metric can be  defined in the super-AdS space
and the bilinear form WZ term can be constructed as follows.
In terms of light-cone indices
$\hat{a}=(+,-,i,i')$ $\hat{i}=(i,i')$ 
and the light cone projection operators for spinors $\lcpm_\pm $
\bea
&&~\theta^\pm=\lcpm_\pm\theta~,~
\zeta_\pm=\zeta\lcpm_\pm,~Q_\pm=Q\lcpm_\pm~,~
\lcpm_\pm \equiv \frac{1}{2}\Gamma_\pm \Gamma_\mp,
~\Gamma_\pm=\frac{1}{\sqrt{2}}(\Gamma_9\pm \Gamma_0)~~,
\eea
the Cartan 1-forms for the super-AdS$_5\times$S$^5$ space 
are written as
\bea
G^{-1}_{\rm AdS}dG_{\rm AdS}
={\bf L}_{\rm AdS}^{+}P_{+}
+{\bf L}_{\rm AdS}^{-}P_{-}
+{\bf L}_{\rm AdS}^{\hat{i}}P_{\hat{i}}
+L_{\rm AdS}^+Q_+ +L_{\rm AdS}^-Q_-
+ {\bf L}_{*\rm AdS}^{\hat{i}}P^*_{\hat{i}}
+\frac{1}{2}{\bf L}_{\rm AdS}^{\hat{i}\hat{j}}M_{\hat{i}\hat{j}}~~~,
\eea 
where 
\bea
P_{i}^*&=&M_{0i},~~~~~~P_{i'}^*=M_{9i'}.
\eea
The MC equation for $Q_-$ is given by 
\bea
dL^-_{\rm AdS}&+&\frac{1}{4}
{\bf L}^{\hat{i}\hat{j}}_{\rm AdS}\Gamma_{\hat{i}\hat{j}}L^-_{\rm AdS}
-\frac{\sqrt{2}}{4}{\bf L}^{\hat{i}}_{*\rm AdS}\Pi\Gamma_{\hat{i}}\Pi
\Gamma_-L^+_{\rm AdS}
\nn\\
&+&\frac{1}{2\sqrt{2}R} \epsilon\left(
-2{\bf L}_{\rm AdS}^+\Pi L_{\rm AdS}^- 
-{\bf L}_{\rm AdS}^{\hat{i}}\Pi\Gamma_{\hat{i}}\Gamma_- 
L_{\rm AdS}^+
\right)
=0~
\label{AdSMCf}
\eea  
with $\Pi=\Gamma_{1234}$ 
and the MC equation for $Q_+$ has a similar form to this.
From these MC equations it can be seen that structure constants
$f_{P_+~Q_-}^{Q_-}$, $f_{P_{\hat{i}}~Q_+}^{Q_-}$,
$f_{P_-~Q_+}^{Q_+}$ and $f_{P_{\hat{i}}~Q_-}^{Q_+}$ are
present and the nondegenerate group metric can be defined consistently.
The bilinear form WZ term can be constructed
 for the super-AdS space \cite{AdSWZ}
\bea
\tilde{B}_{[2],{\rm AdS}}=R\left(
L_{\rm AdS}^{\alpha\alpha' I} 
C_{\alpha\beta}C'_{\alpha'\beta'}\tau_{1,IJ} L_{\rm AdS}^{\beta\beta' J}
-d\theta^{\alpha\alpha' I} 
C_{\alpha\beta}C'_{\alpha'\beta'}\tau_{1,IJ} d\theta^{\beta\beta' J}\right)~~~.
\label{AdSBB}
\eea
Here the nondegenerate metric for spinor index is 
$\rho_{Q_{\alpha\alpha' I} Q_{\beta\beta' J}}=C_{\alpha\beta}C'_{\alpha'\beta'}
(\tau_1)_{IJ}\equiv \rho_{\alpha\beta}$,
and this is used for the WZ term as $L^\alpha L^\beta \rho_{\alpha \beta}$.
It gives $d(L^{\alpha } L^{\beta }\rho_{\alpha \beta})
=L^{\alpha }{\bf L}^c L^{\beta } f_{\alpha c \beta}\sim H_{[3]}$ 
with totally antisymmetric structure constants defined as $f^\gamma_{c \beta}\rho_{\gamma \alpha}$.
In \cite{AdSWZ} it was shown that \bref{AdSBB} gives correct exterior
derivative $d \tilde{B}_{[2]}=H_{[3]}$, $\kappa$-invariance of the total action and 
the correct flat limit.
It was also shown that the second term in \bref{AdSBB} is required 
for the pseudo-supersymmetry invariance 
giving the correct string charge in the superalgebra.

\vskip 3mm

Let us discuss the Penrose limit of this bilinear form WZ term, \bref{AdSBB}. 
The Penrose limit is taken as the following rescaling \cite{HKSpp}
\bea
\theta^+\to \Omega \theta^+,~~\theta^- \to \theta^-
\label{PLtheta}
\eea
in addition to \bref{Penvar}.
This corresponds to $s=\Omega$ in \bref{exps}  
and the scaling dimensions $N_A$ in \bref{mcss} to be
\bea
N_{P_+}=2,~~N_{P_{\hat{i}}}=1,~~N_{P_-}=0,~~N_{P_{\hat{i}}^*}=1,~~
N_{M_{\hat{i}\hat{j}}}=0,~~
N_{Q_+}=1,~~N_{Q_{-}}=0\label{DPMQ} ~~~.
\eea
Under this rescaling the AdS-WZ term \bref{AdSBB} becomes
\bea
\Omega^{2}\tilde{B}_{[2],{\rm AdS}}&=&\frac{iR}{\sqrt{2}}
\left\{
\Omega^2\bar{L}_{(1)}^+\Gamma^+\tau_1 \Pi L^+_{(1)}
-\bar{L}^-_{(0)}\Gamma^-\tau_1 \Pi L^-_{(0)}
-\Omega^2 ~ 2\bar{L}^-_{(0)}\Gamma^-\tau_1 \Pi L^-_{(2)}
\nn\right.\\
&&~~~~\left.
-\Omega^2d\bar{\theta}^{+} \Gamma^+\tau_1 \Pi d\theta^{+}
+d\bar{\theta}^-\Gamma^-\tau_1 \Pi d\theta^-+o(\Omega^3)
\right\}
\eea 
where $\bar{L}=L{\cal C}$ and ${\cal C}$ 
is the charge conjugation matrix in 10 dimensions.
The scaling factor $\Omega^2$ of $\tilde{B}_{[2]}$
is determined by requiring the same scaling weight as the 
string kinetic term, the Nambu-Goto action.

Under the Penrose limit the 
$0$-th order equation of the MC equation for $Q_-$ \bref{IWmc} 
becomes a trivial one according to  \bref{DPMQ}
\bea
dL^-_{(0)}+ f_{M_{\hat{i}\hat{j}} ~Q_-}^{Q_-}
{\bf L}^{\hat{i}\hat{j}}_{(0)}L^-_{(0)}=0~~.
\eea 
This implies that the group metric is degenerate 
in the ``$-$" spinor direction.
In order to make it nondegenerate
$f_{P_+~Q_-}^{Q_-}$ and $f_{P_{\hat{i}}~Q_+}^{Q_-}$ must be 
included, 
so additional term is required 
whose scaling dimension is $N_{P_+}+N_{Q_-}=N_{P_{\hat{i}}}+N_{Q_+}=2$.
It is 
$L^-_{(2)} \equiv \tilde{L}^-_{\rm pp}$ and must be maintained
in the limiting procedure, and
explicit computation confirms that $\tilde{L}_{\rm pp}^-=L^-_{(2)}$ is 
the next to leading term.
Under the Penrose limit the Cartan 1-forms of the super-AdS group
reduce to those of the super-pp background as
\bea
\left\{\begin{array}{ccll}
\Omega^{N_A}L_{\rm AdS}^A&=&\Omega^{N_A}L_{\rm pp}^A+o(\Omega^{N_A+2})&
\cdots {\rm for }~{A\neq Q_-}\\
L_{\rm AdS}^-&=&L_{\rm pp}^-+\Omega^2 \tilde{L}_{\rm pp}^-+o(\Omega^4)
\end{array}\right.~~~
\eea
and they satisfy the following MC equations 
for bosonic and fermionic Cartan 1-forms respectively
\bea
&&\left\{\begin{array}{l}
d{\bf L}_{\rm pp}^+-\frac{1}{\sqrt{2}}{\bf L}_{\rm pp}^{\hat{i}}{\bf L}^{\hat{i}}_{*\rm pp}
-i\bar{L}_{\rm pp}\Gamma^+ L=0
\\
d{\bf L}_{\rm pp}^--i\bar{L}_{\rm pp}\Gamma^- L_{\rm pp}=0
\\
d{\bf L}_{\rm pp}^{\hat{i}}+\frac{1}{\sqrt{2}}{\bf L}_{\rm pp}^-{\bf L}^{\hat{i}}_{*\rm pp}
+{\bf L}_{\rm pp}^{\hat{j}}{\bf L}_{\rm pp}^{\hat{j}\hat{i}}
-i\bar{L}_{\rm pp}\Gamma^{\hat{i}} L_{\rm pp}=0
\\
d{\bf L}^{i}_{*\rm pp}-4\sqrt{2}\mu^2{\bf L}_{\rm pp}^-{\bf L}_{\rm pp}^{i}
+{\bf L}^{j}_{*\rm pp}{\bf L}_{\rm pp}^{ji}
+2\sqrt{2}i\mu \bar{L}_{\rm pp}\Pi\Gamma^{i}\epsilon L_{\rm pp}=0
\\
d{\bf L}^{i'}_{*\rm pp}-4\sqrt{2}\mu^2{\bf L}_{\rm pp}^-{\bf L}_{\rm pp}^{i'}
+{\bf L}^{j'}_{*\rm pp}{\bf L}_{\rm pp}^{j'i'}
+2\sqrt{2}i\mu \bar{L}_{\rm pp}\Pi'\Gamma^{i'}\epsilon L_{\rm pp}=0
\\
d{\bf L}_{\rm pp}^{ij}+{\bf L}_{\rm pp}^{ki}{\bf L}_{\rm pp}^{jk}
+2i\mu \bar{L}_{\rm pp}\Pi\Gamma^{-ij}\epsilon L_{\rm pp}=0
\\
d{\bf L}_{\rm pp}^{i'j'}+{\bf L}_{\rm pp}^{k'i'}{\bf L}_{\rm pp}^{j'k'}
+2i\mu \bar{L}_{\rm pp}\Pi'\Gamma^{-ij}\epsilon L_{\rm pp}=0
\end{array}\right. \label{ppMCb}\\
&&
\left\{\begin{array}{l}
dL_{\rm pp}^+
+\frac{1}{4}(
{\bf L}_{\rm pp}^{\hat{i}\hat{j}}\Gamma_{\hat{i}\hat{j}}L_{\rm pp}^+
+\sqrt{2}{\bf L}^{\hat{i}}_{*\rm pp}\Gamma_{+\hat{i}}L_{\rm pp}^-
)
+\mu \epsilon\left(
2{\bf L}_{\rm pp}^-\Pi L^+ 
-\Pi{\bf L}_{\rm pp}^{\hat{i}}\Gamma_{\hat{i}+}L_{\rm pp}^-
\right)
=0\\
dL^-_{\rm pp}+\frac{1}{4}
{\bf L}^{\hat{i}\hat{j}}_{\rm pp}\Gamma_{\hat{i}\hat{j}}L^-_{\rm pp}=0\\
d\tilde{L}^-_{\rm pp}+\frac{1}{4}
({\bf L}^{\hat{i}\hat{j}}_{\rm pp}\Gamma_{\hat{i}\hat{j}}\tilde{L}^-_{\rm pp}
-\sqrt{2}{\bf L}^{\hat{i}}_{*\rm pp}\Pi\Gamma_{\hat{i}}\Pi
\Gamma_-L^+_{\rm pp})
+\mu \epsilon\left(
-2{\bf L}_{\rm pp}^+\Pi L_{\rm pp}^- 
-{\bf L}_{\rm pp}^{\hat{i}}\Pi\Gamma_{\hat{i}}\Gamma_- 
L_{\rm pp}^+
\right)
=0
~~~,\label{tildemc}
\end{array}\right. \label{ppMC}
\eea
where $\Pi'=\Gamma_{5678}$.
These MC equations coincide with the ones obtained by the direct computation
\cite{Me1} using
\bea
G^{-1}_{\rm pp}dG_{\rm pp}
={\bf L}_{\rm pp}^{+}P_{+}
+{\bf L}_{\rm pp}^{-}P_{-}
+{\bf L}_{\rm pp}^{\hat{i}}P_{\hat{i}}
+L_{\rm pp}^+Q_+ +L_{\rm pp}^-Q_-
+ {\bf L}_{*\rm pp}^{\hat{i}}P^*_{\hat{i}}
+\frac{1}{2}{\bf L}_{\rm pp}^{\hat{i}\hat{j}}M_{\hat{i}\hat{j}}~~~,
\eea
except for the last equation for $\tilde{L}^-_{\rm pp}$.
It would be obtained
if one uses an extended super-pp algebra with a fermionic center.
The last equation of \bref{ppMC} has  the same form 
as the MC equation of the AdS \bref{AdSMCf} 
then it becomes nondegenerate. 

Following to above 
limiting procedure from the super-AdS WZ term \bref{AdSBB} 
we propose  a superstring action in the super-pp-wave background
as
\bea
S&=&\displaystyle\int d^2\sigma~{\cal L}
=\displaystyle\int d^2\sigma\left({\cal L}_0+{\cal L}_{WZ}\right)
\nn\\
{\cal L}_0&=&-~T\sqrt{-h}h^{\uu\vv}~{\bf L}_{{\rm }\uu}^{~~\hat{a}}
{\bf L}_{{\rm }\vv}^{~~\hat{b}}\eta_{\hat{a}\hat{b}}~~,~~
{\cal L}_{WZ}
=~T \tilde{B}_{[2],{\rm pp}}
~~~~\label{action}
\eea
where the WZ term is 
\bea
\tilde{B}_{[2],{\rm pp}}=\frac{i}{4\mu}
\left(
\bar{L}_{\rm pp}^{+I}\Gamma_-(\tau_1)_{IJ}\Pi L_{\rm pp}^{+J}
-2\bar{L}_{\rm pp}^{-I}\Gamma_+(\tau_1)_{IJ} \Pi\tilde{L}_{\rm pp}^{-J}
-d \bar{\theta}^{+I}\Gamma_-(\tau_1)_{IJ}\Pi d \theta^{+J}
\right)\label{WZppnew}~~~
\eea 
and $\uu, \vv=\tau,\sigma=0,1$ are the world volume indices
and Cartan 1-forms are $L^A=dz^M L_M^{~~A}=d\sigma^\uu L_\uu^{~~A}$
with $z^M=(x^m, \theta^\mu)$.\par\vskip 6mm

It satisfies 
the following criteria: \par
{\bf (i)}   giving the correct three form, 
$d\tilde{B}_{[2]}=H_{[3]}$ for $dH_{[3]}=0$,\par
{\bf (ii)}  $\kappa$-invariance of the total action. \par 
\vskip 6mm
\noindent
{\bf (i) Three form $H_{[3]}$}\par
\indent
From now on the currents $L_{\rm pp}^A$ is denoted just as 
$L^A$.
The first condition {\bf (i)} is checked 
by using the MC equations \bref{ppMCb} and \bref{ppMC},
\bea
d\tilde{B}_{[2]{\rm }}&=&\frac{i}{4\mu}
\left(
2\bar{L}^{+I}\Gamma_-(\tau_1)_{IJ}\Pi dL^{+J}
-2d\bar{L}^{-I}\Gamma_+(\tau_1)_{IJ} \Pi\tilde{L}^{-J}
-2\bar{L}^{-I}\Gamma_+(\tau_1)_{IJ} \Pi d\tilde{L}^{-J}
\right)\nn\\
&=&
-i (\bar{L}^+{\bf L}^-\Gamma_-\tau_3 L^+
+\bar{L}^+{\bf L}^{\hat{i}}\Gamma_{\hat{i}}\tau_3 L^-
)
+i\frac{1}{2\sqrt{2}\mu}\bar{L}^+(-{\bf L}^{i}_{*}\Gamma_{i}+
{\bf L}^{i'}_{*}\Gamma_{i'})\Pi\tau_1
 L^-\nn\\
&&
-i (\bar{L}^-{\bf L}^+\Gamma_-\tau_3 L^-
+\bar{L}^-{\bf L}^{\hat{i}}\Gamma_{\hat{i}}\tau_3 L^+
)
+i\frac{1}{2\sqrt{2}\mu}\bar{L}^+({\bf L}^{i}_{*}\Gamma_{i}-
{\bf L}^{i'}_{*}\Gamma_{i'})\Pi\tau_1
 L^-\nn\\
&=&-i
\bar{L}
{\bf L}^{\hat{a}}\Gamma_{\hat{a}}\tau_3 
L
\nn\\
&=&H_{[3]}~~~.\label{H3}
\eea

\noindent
{\bf (ii) $\kappa$-invariance}\par
\indent
The second condition {\bf (ii)} $\kappa$-invariance 
of the action is confirmed as follows.
Let us denote an arbitrary variation of the coset element $\delta G$
as the following combination 
\bea
G^{-1}\delta G=\Delta L^A T_A =\delta z^M L_M^{~~A}T_A \label{DeltaL}~~~,
\eea
and the variation of the Cartan 1-form is given by
\bea
\delta L^A T_A=\delta (G^{-1}dG)=\left\{
d(\Delta L^A)-\Delta L^B L^C f_{BC}^A
\right\}T_A ~~~\label{DelradL}.
\eea
Then an arbitrary variation of a Cartan 1-form is
obtained by referring to the corresponding MC equation.

The $\kappa$-variations ($\delta\theta^\mu =\kappa^\mu$) of Cartan 1-forms 
are given by
\bea
\delta_\kappa {\bf L}^{\hat{a}}&=&-f_{BC}^{\hat{a}} \Delta_\kappa L^B L^C\nn\\
\delta_\kappa L^{\alpha}&=&d(\Delta_\kappa L^\alpha)
-f_{BC}^{\alpha} \Delta_\kappa L^B L^C
\eea 
where $\Delta_\kappa {\bf L}^{\hat{a}}=0$ is imposed for the 
$\kappa$-variations.
Concrete expression is
\bea
&&\left\{\begin{array}{l}
\delta_\kappa {\bf L}^+-\frac{1}{\sqrt{2}}
(\Delta_\kappa {\bf L}^{\hat{i}}{\bf L}_*^{\hat{i}}
+{\bf L}^{\hat{i}}\Delta_\kappa{\bf L}_*^{\hat{i}})
-2i\bar{L}\Gamma^+\Delta_\kappa L=0\\
\delta_\kappa {\bf L}^--2i\bar{L}\Gamma^-\Delta_\kappa L=0\\
\delta_\kappa {\bf L}^{\hat{i}}+\frac{1}{\sqrt{2}}
(\Delta_\kappa {\bf L}^-{\bf L}_*^{\hat{i}}
+{\bf L}^-\Delta_\kappa{\bf L}_*^{\hat{i}})
+(\Delta_\kappa {\bf L}^{\hat{j}}{\bf L}^{\hat{j}\hat{i}}
+{\bf L}^{\hat{j}}\Delta_\kappa {\bf L}^{\hat{j}\hat{i}}
)
-2i\bar{L}\Gamma^{\hat{i}}\Delta_\kappa L=0\\
\end{array}\right.\label{bosDel}\\
&&\left\{\begin{array}{l}
\delta_\kappa L^++\mu\epsilon
\left\{
2(\Delta_\kappa{\bf L}^-\Pi L^++{\bf L}^-\Pi \Delta_\kappa L^+)
-\Pi( \Delta_\kappa {\bf L}^{\hat{i}}\Gamma_{\hat{i}}\Gamma_+L^-
+{\bf L}^{\hat{i}}\Gamma_{\hat{i}}\Gamma_+ \Delta_\kappa L^-)
\right\}\\
~~+\frac{1}{4}( \Delta_\kappa{\bf L}^{\hat{i}\hat{j}}\Gamma_{\hat{i}\hat{j}}L^+
+{\bf L}^{\hat{i}\hat{j}}\Gamma_{\hat{i}\hat{j}}\Delta_\kappa L^+)
-\frac{1}{2\sqrt{2}}(
\Delta_\kappa{\bf L}^{\hat{i}}_* \Gamma_{\hat{i}+}L^-
+{\bf L}^{\hat{i}}_* \Gamma_{\hat{i}+}\Delta_\kappa L^-)=0\\
\delta_\kappa L^-
+\frac{1}{4}( \Delta_\kappa{\bf L}^{\hat{i}\hat{j}}\Gamma_{\hat{i}\hat{j}}L^-
+{\bf L}^{\hat{i}\hat{j}}\Gamma_{\hat{i}\hat{j}}\Delta_\kappa L^-)=0\\
\delta_\kappa \tilde{L}^- -\mu\epsilon
\left\{
2(\Delta_\kappa{\bf L}^+\Pi L^- +{\bf L}^+\Pi \Delta_\kappa L^-)
-\Pi( \Delta_\kappa {\bf L}^{\hat{i}}\Gamma_{\hat{i}}\Gamma_-L^+
+{\bf L}^{\hat{i}}\Gamma_{\hat{i}}\Gamma_- \Delta_\kappa L^+)
\right\}\\
~~+\frac{1}{4}( \Delta_\kappa{\bf L}^{\hat{i}\hat{j}}\Gamma_{\hat{i}\hat{j}}\tilde{L}^-
+{\bf L}^{\hat{i}\hat{j}}\Gamma_{\hat{i}\hat{j}}\Delta_\kappa \tilde{L}^-)
-\frac{1}{2\sqrt{2}}\Pi(
\Delta_\kappa{\bf L}^{\hat{i}}_* \Gamma_{\hat{i}-}\Pi L^+
+{\bf L}^{\hat{i}}_* \Gamma_{\hat{i}-}\Pi\Delta_\kappa L^+)=0~~~~.
\end{array}\right.\label{ferDel}
\eea
Using these relations, the $\kappa$-variation of the total action 
is calculated as
\bea
\delta_\kappa({\cal L}_0+b{\cal L}_{WZ})&=&
-~T\delta_\kappa(
\sqrt{-g}g^{\uu\vv}~{\bf L}_\uu^{\hat{a}}{\bf L}_{\hat{a},\vv})
+bT\epsilon^{\uu\vv}\tilde{B}_{[2]\uu\vv}\nn\\
&=&-2iT
(\overline{\Delta_\kappa L})~(\sqrt{-\det G_{\uu\vv}} G^{\uu\vv} \slbL_\uu {\bf 1}
+b\epsilon^{\uu\vv}\slbL_\uu \tau_3)~L_\vv \nn\\
&=&-2iT(\overline{\Delta_\kappa L})(-\Gamma_{(1)}+b)\epsilon^{\uu\vv}
\slbL_\uu \tau_3~L_\vv
\label{kappa}
\eea
with
\bea
\Gamma_{(1)}=\frac{1}{2\sqrt{-\det G_{\uu\vv}}}\tau_3\slbL~,~
\Gamma_{(1)}^2=1~,~{\rm tr} \Gamma_{(1)}=0~,~
-\sqrt{-\det G_{\uu\vv}}G^{\uu\vv}\slbL_\vv=
\Gamma_{(1)}\tau_3\epsilon^{\uu\vv}\slbL_\vv~~.\nn\\
\eea
The action has the $\kappa$ invariance 
whose parameter $\kappa$ is satisfying 
\bea
(-\Gamma_{(1)}+b)\kappa=0~~,~~b=\pm 1~~~.\label{bpm1}
\eea

\par

\subsection{ Gauge invariant action for a super-pp-string}\par
\indent

We present an explicit expression of the gauge invariant action 
for a superstring in the RR pp-wave background, with $b=1$ of \bref{bpm1},
\bea
S&=&\displaystyle\int d^2\sigma~{\cal L}
=\displaystyle\int d^2\sigma\left({\cal L}_0+{\cal L}_{WZ}\right)
~~,~~
{\cal L}_0=-~T
\sqrt{-h}h^{\uu\vv}~{\bf L}_{{\rm }\uu}^{~~\hat{a}}{\bf L}_{{\rm }\vv}^{~~\hat{b}}\eta_{\hat{a}\hat{b}}\nn\\
{\cal L}_{WZ}&=&
~T
\epsilon^{\uu\vv}\frac{i}{4\mu}
\left(
-\bar{L}_{{\rm }\uu}^{~~+}\Gamma_-\tau_1\Pi L_{{\rm }\vv}^{~~+}
+2\bar{L}_{{\rm }\uu}^{~~-}\Gamma_+\tau_1 \Pi\tilde{L}_{{\rm }\vv}^{~~-}
+\partial_\uu \bar{\theta}^{+}\Gamma_-\tau_1\Pi\partial_\vv \theta^{+}
\right)~~~~~\label{Action}
\eea
where the Cartan 1-forms in the pp-wave background are 
obtained from the ones in the super-AdS$_5\times$S$^5$ given
in the appendix. The results are 
\bea
\left\{
\begin{array}{lcl}
{\bf L}^+_{\rm }&=&e^+ +
i\bar{\theta}^{+}\Gamma^+\left(
\displaystyle\frac{\sin \frac{\Psi_{+}}{2}}{\frac{\Psi_{+}}{2}}\right)^2
(D\theta)^+
+i\bar{\theta}^+\Gamma^+
\left(\displaystyle\frac{\sin \frac{\Psi}{2}}{\frac{\Psi}{2}}\right)^2 \left.(D\theta)^{-}\right|_{\Omega^0}
\\
{\bf L}^-_{\rm }&=&e^- +
i\bar{\theta}^{-}\Gamma^-\left(
\displaystyle\frac{\sin \frac{\Psi_{-}}{2}}{\frac{\Psi_{-}}{2}}\right)^2
\left.(D\theta)^{-}\right|_{\Omega^0}\\
{\bf L}^{\hat{i}}_{\rm }&=&e^{\hat{i}} 
+i\bar{\theta}^{+}\Gamma^{\hat{i}}
\left(\displaystyle\frac{\sin \frac{\Psi_{-}}{2}}{\frac{\Psi_{-}}{2}}\right)^2
\left.(D\theta)^{-}\right|_{\Omega^0}
+i\bar{\theta}^{-}\Gamma^{\hat{i}}
\left(\displaystyle\frac{\sin \frac{\Psi_{+}}{2}}{\frac{\Psi_{+}}{2}}\right)^2
(D\theta)^+\\&&
+i\bar{\theta}^{-}\Gamma^{\hat{i}}
\left.\left(\displaystyle\frac{\sin \frac{\Psi}{2}}{\frac{\Psi}{2}}\right)^2\right|_{\Omega}
 \left.(D\theta)^{-}\right|_{\Omega^0}
\\
L^{+}_{\rm }&=&\displaystyle
\frac{\sin \Psi_{+}}{\Psi_{+}}(D\theta)^{+}
+\left(\left.\lcpm_+\displaystyle\frac{\sin \Psi}{\Psi}\lcpm_-\right)
\right|_{\Omega}
\left.(D\theta)^{-}\right|_{\Omega^0}
\\
L^{-}_{\rm }&=&
\displaystyle\frac{\sin \Psi_{-}}{\Psi_{-}}
\left.(D\theta)^{-}\right|_{\Omega^0}
\\
\tilde{L}^-_{\rm }&=&
\left(
\displaystyle\frac{\sin \Psi_-}{\Psi_-}\right)
\left. (D\theta)^{-}\right|_{\Omega^2}
+\left.\left(
\displaystyle\frac{\sin \Psi_-}{\Psi_-}\right)
\right|_{\Omega^2}\left.(D\theta)^{-}\right|_{\Omega^0}
+\left.\left(\lcpm_-
\displaystyle\frac{\sin \Psi}{\Psi}\lcpm_+\right)
\right|_{\Omega}
(D\theta)^{+}
\end{array}\right.\nn\\
&&\label{C1form}
\eea
with the covariant derivatives 
\bea
&&
\left\{\begin{array}{ccl}
\left. (D\theta)^{-}\right|_{\Omega^0}&=&
d\theta^-
+\frac{1}{4}
\omega^{\hat{i}\hat{j}}\Gamma_{\hat{i}\hat{j}}\theta^-\\
\left. (D\theta)^{-}\right|_{\Omega^2}&=&
-\mu\epsilon\left(
2e^+\Pi \theta^- 
-\Pi e^{\hat{i}}\Gamma_{\hat{i}}\Gamma_-\theta^+
\right)
-\frac{\sqrt{2}}{4}e_*^{\hat{i}}\Pi\Gamma_{\hat{i}}\Pi\Gamma_-\theta^+\\
(D\theta)^{+}&=&d\theta^+
+\frac{1}{4}
\omega^{\hat{i}\hat{j}}\Gamma_{\hat{i}\hat{j}}\theta^+
+\mu\epsilon\left(
2e^-\Pi \theta^+
-\Pi e^{\hat{i}}\Gamma_{\hat{i}}\Gamma_+\theta^-
\right)
-\frac{\sqrt{2}}{4}e_*^{\hat{i}}\Gamma_{\hat{i}}
\Gamma_+\theta^-
\end{array}\right.\label{dthe}
\eea
and arguments of the sin's are
\bea
&&\Psi^2=2\mu i\left[
\epsilon\Pi(2\theta^+ \bar{\theta}^-\Gamma_+
-2\theta^-\bar{\theta}^+\Gamma_-+
\Gamma_+\Gamma^{\hat{i}}\theta^-\bar{\theta}\Gamma_{\hat{i}}
-\Gamma_-\Gamma^{\hat{i}}\theta^+\bar{\theta}\Gamma_{\hat{i}}
)\right.
\label{psipsi}\\
&&~~~~~~~+\left(\Gamma_+
(\Gamma^{\hat{i}}\theta^-\bar{\theta}^-\Gamma_{\hat{i}}
\Pi
-\Gamma^{\hat{i}}\theta^-\bar{\theta}^+\Pi\Gamma_{\hat{i}}
)
+\Gamma_-(
\Gamma^{\hat{i}}\theta^+\bar{\theta}^-\Pi\Gamma_{\hat{i}}
-\Gamma^{\hat{i}}\theta^+\bar{\theta}^+\Gamma_{\hat{i}}\Pi
)\right)\epsilon\nn\\
&&~~~~~~~\left.-\frac{1}{2}
\left(
(\Gamma^{ij}\theta\bar{\theta}^-\Gamma_{ij}
-\Gamma^{i'j'}\theta\bar{\theta}^-\Gamma_{i'j'}
)\Gamma_+
-(\Gamma^{ij}\theta\bar{\theta}^+\Gamma_{ij}
-\Gamma^{i'j'}\theta\bar{\theta}^+\Gamma_{i'j'}
)\Gamma_-
\right)\Pi\epsilon
\right]~~~,\nn\\
&&\left\{\begin{array}{ccl}
\lcpm_+ \Psi^2 \lcpm_+\equiv \Psi_{+}^2&=&2\mu  i
\left[\epsilon\Pi
\Gamma_+\Gamma^{\hat{i}}\theta^-\bar{\theta}^-\Gamma_{\hat{i}}
+\Gamma_+\Gamma^{\hat{i}}\theta^-\bar{\theta}^-\Gamma_{\hat{i}}
\Pi\epsilon\right]\nn\\
\lcpm_- \Psi^2 \lcpm_-\equiv \Psi_{-}^2&=&-\mu {i}
(\Gamma^{ij}\theta^-\bar{\theta}^-\Gamma_{ij}
-\Gamma^{i'j'}\theta^-\bar{\theta}^-\Gamma_{i'j'}
)\Gamma_+\Pi\epsilon\\
\left.\lcpm_+ \Psi^2 \lcpm_-\right|_{\Omega}&=&2\mu i
\left[
\epsilon \Pi(2\theta^+\bar{\theta}^-\Gamma_+
+\Gamma_+\Gamma^{\hat{i}}\theta^-\bar{\theta}^+\Gamma_{\hat{i}}
-\Gamma_+\Gamma^{\hat{i}}\theta^- \bar{\theta}^+\Pi\Gamma_{\hat{i}}\epsilon
\right.
\\
&&\left.-\frac{1}{2}(\Gamma^{ij}\theta^+\bar{\theta}^-\Gamma_{ij}
-\Gamma^{i'j'}\theta^+\bar{\theta}^-\Gamma_{i'j'}
)\Gamma_+\Pi\epsilon\right]\\
\left.\lcpm_- \Psi^2 \lcpm_+\right|_{\Omega}&=&2\mu i
\left[
\epsilon \Pi(-2\theta^-\bar{\theta}^+\Gamma_-
-\Gamma_-\Gamma^{\hat{i}}\theta^+\bar{\theta}^-\Gamma_{\hat{i}}
+\Gamma_-\Gamma^{\hat{i}}\theta^+ \bar{\theta}^-\Pi\Gamma_{\hat{i}}\epsilon
\right.
\\
&&\left.+\frac{1}{2}(\Gamma^{ij}\theta^-\bar{\theta}^+\Gamma_{ij}
-\Gamma^{i'j'}\theta^-\bar{\theta}^+\Gamma_{i'j'}
)\Gamma_-\Pi\epsilon\right]\\
\left.\lcpm_- \Psi^2 \lcpm_- \right|_{\Omega^2}&=&
2\mu i\left[
\epsilon\Pi\Gamma_-\Gamma^{\hat{i}}\theta^+\bar{\theta}^+\Gamma_{\hat{i}}
-\Gamma_-\Gamma^{\hat{i}}\theta^+\bar{\theta}^+\Gamma_{\hat{i}}\Pi\epsilon
\right]
~~.\end{array}\right.
\eea
The Cartan 1-forms are same as the one in \cite{Me1} 
except for $\tilde{L}^-$. 
From now on we use the ``$x$-coordinates" in \bref{yandx} 
for simplicity
in which the bosonic Cartan 1-forms are given by
\bea
\left\{\begin{array}{lcl}
 e^+&=&dx^+
 -2\mu^2(x^{\hat{i}})^2 dx^-\\
e^{-}&=&dx^-\\
e^{\hat{i}}&=&dx^{\hat{i}}\\
e^{\hat{i}}_*&=&-4\sqrt{2}\mu^2 x^{\hat{i}} dx^-\\
\omega^{\hat{i}\hat{j}}&=&0
\end{array}\right.\label{ppex}
\eea

We confirm that this action reduces to following known ones \par
{\bf (i)} light-cone action in the light cone gauge,\par
{\bf (ii)} flat action in the flat limit $\mu\to 0$.\par
\vskip 6mm
\noindent
{\bf (i)  Light-cone gauge action}\par
\indent
In the light-cone gauge with the conformal metric
\bea
x_+=p_+\tau~~,~~\theta^-=0\label{lcgg}~~~,~h^{\uu\vv}=\eta^{\uu\vv}
\eea
the Cartan 1-forms reduce into 
\bea
\left\{
\begin{array}{lcl}
{\bf L}^+_{\rm }&=&e^+ +
i\bar{\theta}^{+}\Gamma^+
(d\theta^+ +
2\mu\epsilon e_+\Pi\theta^+ )
\\
{\bf L}^-_{\rm }&=&e_+ \\
{\bf L}^{\hat{i}}_{\rm }&=&e^{\hat{i}} 
\\
L^{+}_{\rm }&=&d\theta^++2\mu\epsilon e_+\Pi\theta^+
\\
L^{-}_{\rm }&=&0\\
\tilde{L}^-_{\rm }&=&
\mu\epsilon \Pi e^{\hat{i}}\Gamma_{\hat{i}-}\theta^+
-\frac{\sqrt{2}}{4}e_*^{\hat{i}}\Pi\Gamma_{\hat{i}}\Pi
\Gamma_-\theta^+~~~.
\end{array}\right.\nn\\
&&\label{C1formlc}
\eea
The light-cone action of a superstring in the super-pp-wave background 
is given by
\bea
&&S=-T\displaystyle\int~d^2\sigma~
\left[\frac{1}{2}\left(\partial^\uu x^{\hat{i}}\partial_\uu x^{\hat{i}}
+m^2 (x^{\hat{i}})^2\right)
-ip_+\left(\bar{\theta}^+\Gamma^+(\partial_0+\tau_3 \partial_1)\theta^+
+m\bar{\theta}^+\Gamma^+\epsilon \Pi\theta^+
\right)
\right]~~,\nn\\
&&~~~\hskip 90mm~~~~~~m\equiv 2\mu p_+~~~
\eea
which coincides with the one obtained by Metsaev \cite{Me1}.\par\vskip 6mm\noindent
{\bf (ii) Flat action}\par
\indent
The flat limit is realized by taking $\mu \to 0$ limit.
The kinetic term  ${\cal L}_0$ in the flat limit 
is straightforwardly obtained by taking $\mu \to 0$ 
in Cartan 1-forms ${\bf L}^{\hat{a}}$ as
\bea
{\bf L}^{\hat{a}}&=&dx^{\hat{a}}+i\bar{\theta}\Gamma^{\hat{a}}d\theta~~
.
\eea
The WZ term ${\cal L}_{WZ}$ in the flat limit
must be evaluated with care.
Term proportional to $1/\mu$ is absent in \bref{action}.
Terms proportional to $\mu^0$ are obtained from terms proportional to $\mu^1$ 
in $L$'s 
\bea
L^+&=&d\theta^+ 
+\mu\left[
\epsilon\Pi(2e^-\theta^+ -e^{\hat{i}}\Gamma_{\hat{i}}\Gamma_+\theta^-)
+\frac{1}{3!}(\Psi_+^{~2}d\theta^+ +\lcpm_+\Psi^2\lcpm_- d\theta^-)
\right]+o(\mu^2)\nn\\
\tilde{L}^-&=&\mu\left[
-\epsilon\Pi(2e^+\theta^- -e^{\hat{i}}\Gamma_{\hat{i}}\Gamma_-\theta^+)
+\frac{1}{3!}(\Psi_-^{~2}d\theta^- 
+\left.\lcpm_-\Psi^2\lcpm_-\right|_{\Omega^2} d\theta^-
+\lcpm_-\Psi^2\lcpm_+ d\theta^+
)
\right]+o(\mu^2)~~~.\nn\\
L^-&=&d\theta^-+o(\mu^2)
\eea 
The WZ term becomes 
\bea
{\cal L}_{WZ}&=&
~\left.T\frac{i}{4\mu}
\mu\left(2d\bar{\theta}^+\Gamma_-\tau_1\Pi 
\left.(L^+)\right|_{\mu^1}
-2d\bar{\theta}^{-}\Gamma_+\tau_1 \Pi
\left.(\tilde{L}^-)\right|_{\mu^1}
+o(\mu^3)
\right)\right|_{\mu \to 0}~~~~~\nn\\
&=&
T\left[
idx^{\hat{a}} \bar{\theta}\Gamma_{\hat{a}}d \theta
+\frac{1}{3}\bar{\theta}\Gamma^{\hat{a}}\tau_3 d \theta~
\bar{\theta}\Gamma^{\hat{a}} d\theta
+\frac{1}{3}\bar{\theta}\Pi\Gamma^{\hat{i}}\tau_1 d \theta~
\bar{\theta}\Pi\Gamma^{\hat{i}}\epsilon d\theta\right.\nn\\
&&\left.-\frac{1}{12}\left(
\bar{\theta}^-\Pi\Gamma^{-{i}{j}}\tau_1 d \theta~
\bar{\theta}\Pi\Gamma^{-{i}{j}}\epsilon d\theta
-
\bar{\theta}^-\Pi\Gamma^{-{i'}{j'}}\tau_1 d \theta~
\bar{\theta}\Pi\Gamma^{-{i'}{j'}}\epsilon d\theta
\right)
\right]\nn\\
&&=T\left[
idx^{\hat{a}} \bar{\theta}\Gamma_{\hat{a}}d \theta
+\frac{1}{2}\bar{\theta}\Gamma^{\hat{a}}\tau_3 d \theta~
\bar{\theta}\Gamma^{\hat{a}} d\theta\right]
\eea
where the last equality is derived by using the relation 
in the section ${\bf 2.3}$ of \cite{AdSWZ}.
This action is the Green-Schwarz superstring action in a flat space 
\cite{GS}.

\par

\section{ Hamiltonian and equations of motion}\par
\indent

In this section we calculate the Hamiltonian, constraints and
equations of motion of the system.  
Before examining the superstring in the pp-wave background,
particle, superparticle and  bosonic string cases are examined.

\subsection{ Bosonic particle}\par
\indent

The action for a bosonic particle system in a curved background is 
\bea
S_{\rm PA}&=&\displaystyle\int d\tau
~\frac{1}{2e}{\dot{x}}^mg_{mn}{\dot{x}}^n\nn
~~.
\eea
The canonical momentum of $x^m$ is
$p_m=\delta S_{PA}/\delta \dot{x}^m=(1/e)g_{mn}\dot{x}^n$ and 
the gauge invariant Hamiltonian is
\bea
{\cal H}_{\rm PA}&=&e H_{\rm PA} ~~,~~H_{\rm PA}=\frac{1}{2}p_mg^{mn}p_n=0~~.
\eea
It leads to the following equations of motion in $e=1$ gauge
\bea
\left\{\begin{array}{l}
\dot{x}^m=g^{mn}p_n\\
\dot{p}_m=-\frac{1}{2}(\partial_m g^{nl})p_np_l
\end{array}\right.
\eea
then
\bea
\ddot{x}^m=-{\bf \Gamma}^m_{~nl}\dot{x}^n\dot{x}^l
~~~
\eea
with the Affine connection coefficients defined by
${\bf \Gamma}^l_{~nk} =\frac{1}{2}g^{lm}
(-\partial_m g_{nk}+\partial_{(n}g_{k)m})$.

For the pp-wave background the metric is given by
\bea
g_{mn}=\left(
\begin{array}{ccc}
g_{++}&g_{+-}&g_{+{\hat{j}}}\\
g_{-+}&g_{--}&g_{-{\hat{j}}}\\
g_{{\hat{i}}+}&g_{{\hat{i}}-}&g_{{\hat{i}}{\hat{j}}}
\end{array}
\right)=
\left(
\begin{array}{ccc}
0&1&0\\
1&-4\mu^2 \bmx^2&0\\
0&0&\delta_{{\hat{i}}{\hat{j}}}\end{array}
\right)~,~
g^{mn}=\left(
\begin{array}{ccc}
4\mu^2 \bmx^2&1&0\\
1&0&0\\
0&0&\delta_{{\hat{i}}{\hat{j}}}\end{array}
\right)~
~,\label{metricg}
\eea
the Hamiltonian in the $e=1$ gauge is given by
\bea
{\cal H}_{\rm PA}&=&\frac{1}{2}\left[
2p_+p_- +{\bmp}^2 +(2\mu p_+)^2 {\bmx}^2
\right]
\eea
and equations of motion are 
\bea
\left\{\begin{array}{l}
\dot{x}^+=p_-+(2\mu)^2 {\bmx}^2p_+\\
\dot{x}^-=p_+\\
\dot{\bmx}={\bmp}
\end{array}\right.~~,~~
\left\{\begin{array}{l}
\dot{p}_-=0\\
\dot{p}_+=0\\
\dot{\bmp}=-(2\mu p_+)^2{\bmx}
\end{array}\right.\label{eqPA}
\eea
with $\mbox{\boldmath $x$}=x^{\hat{i}}$ and ${\bmp}=p_{\hat{i}}$.
The second order equations of motion can be written as
\bea
\left\{\begin{array}{l}
\ddot{x}^+=(2\mu)^2p_+\dot{({\bmx}^2)}\\
\ddot{x}^-=0\\
\ddot{\bmx}=-\omega^2 {\bmx} ~~,~~\omega=2\mu p_+
\end{array}\right.
\label{harm}
\eea 
and \bref{eqPA} and \bref{harm} are solved as
\bea
\left\{\begin{array}{l}
x^+=\left(p_- +2\mu^2 p_+ ({\bma}^2+\tilde{\bma}^2)\right)\tau
-\frac{1}{2}\mu({\bma}^2-\tilde{\bma}^2) 
\sin (4\mu x^-)
-\mu{\bma}\cdot\tilde{\bma} \left( \cos(4\mu x^-)-1 \right)  +x_0^+
\\
x^-=p_+\tau+x^-_0\\
{\bmx}={\bma} \sin (2\mu x^-)+\tilde{\bma} \cos (2\mu x^-)
\end{array}\right. \label{solPA}~~,
\eea
where constants \mbox{\boldmath $\alpha$} and 
\mbox{\boldmath $\tilde\alpha$} are 
amplitudes of harmonic motions in 
transverse directions.
The Hamiltonian is also written as
\bea
{\cal H}_{\rm PA}&=&\frac{1}{2}\left[
2p_+p_- +\omega^2({\bma}^2+\tilde{\bma}^2)\right]
\eea

The canonical momenta $p_{\hat i}$ are not constants of motion but the
constants of motion are the  
global translation and boost charges $P_{\hat{a}}$ and $P_{\hat{i}}^*$. 
Their forms in terms of the canonical variables are calculated  from
\bea
\lambda^{\hat{a}}P_{\hat{a}}=p_m ~\delta_\lambda x^m
=p_m~\Delta_\lambda L^A~(L^{-1})_A^{~~m}
~~,~~G_{\rm pp}^{-1}\delta_\lambda G_{\rm pp}
=G_{\rm pp}^{-1}(\lambda^{\hat{a}}P_{\hat{a}}) G_{\rm pp}
=\Delta_\lambda L^A ~T_A~~,\nn\\
\eea
etc. and are 
\bea
P_{\hat{a}}&=&
\left\{\begin{array}{l}
{P}_-=p_-\\
{P}_+=p_+\\
{P}_{\hat{i}}=
{p_{\hat{i}}}\cos (2\mu x^-)+2\mu p_+ {x_{\hat{i}}}\sin (2\mu x^-)
=\omega \alpha_{\hat{i}}~
\end{array}\right.\\
{P}_{\hat{i}}^*&=&
(-\frac{1}{2\sqrt{2}\mu})
 \bigl({p_{\hat{i}}}\sin (2\mu x^-)
+2\mu p_+ {x_{\hat{i}}}\cos (2\mu x^-) \bigr) 
=(-\frac{1}{2\sqrt{2}\mu})\omega \tilde{\alpha}_{\hat{i}}~\nn~.
\eea
In terms of global charges the Hamiltonian can be written as
the quadratic Casimir operator of the pp-algebra
\bea
{\cal H}_{\rm PA}&=&\frac{1}{2}(
{P}_{\hat{a}}\eta^{\hat{a}\hat{b}}{P}_{\hat{b}}
+P_{\hat{i}}^*\eta^{\hat{i}\hat{j}}_*P_{\hat{j}}^*)
~~,~~
\eta^{\hat{i}\hat{j}}_*\equiv \delta^{\hat{i}\hat{j}}(2\sqrt{2}\mu)^2~~.
\label{PPppstar}\label{hamPA}
\eea
This Hamiltonian is justified algebraically 
as it commutes with all global space-time symmetry
charges. 
It is also obtained by the Penrose limit of 
the Hamiltonian of the AdS particle which is also the
quadratic Casimir operator of the AdS algebra;
\bea
&&
\left\{\begin{array}{l}
c_{\rm AdS}=\displaystyle\sum
P_{{a}} P_{{b}}\eta^{{a}{b}}+
\displaystyle\sum(M_{{a}{b}})^2\\
~~~~~~~^{{a}=0,1,\cdot,4}~~~~~~~~
^{{a}{b}=0,\cdot,4~} \\
c_{\rm S}=\displaystyle\sum
P_{{a'}} P_{{b'}}\eta^{{a'}{b'}}+
\displaystyle\sum(M_{{a'}{b'}})^2\\
~~~~~~~^{{a'}=5,\cdot,9}~~~~~~~~
^{{a'}{b'}=5,\cdot,9}
\end{array}\right. \nn\\
&&~~~
\stackrel{\rm Penrose~limit}{\longrightarrow}
\left\{\begin{array}{l}
c_{{\rm pp}(-2)}=\displaystyle\sum
P_{\hat{a}}P_{\hat{b}}\eta^{\hat{a}\hat{b}}
+\displaystyle\sum 
 {P_{\hat{i}}^*}
{P_{\hat{j}}^*}\eta_*^{\hat{i}\hat{j}}
\Rightarrow 2{\cal H}_{\rm PA}~\\
~~~~~~~~~~~~~^{\hat{a}=0,1,\cdot,9}~~~~~~~~~~~~~~
^{\hat{i}=1,\cdots,8}\\
c_{{\rm pp}(-4)}=P_+^2~~.
\end{array}\right.
\eea
There are two quadratic Casimir operators each for the $AdS_5$ and $S^5$.
They turn to become two quadratic Casimir
operators in the pp algebra. The one, which contains the Lorentz 
invariant mass term in a flat limit $\mu\to 0$, 
is the Hamiltonian of this system.
  
\subsection{ Superparticle}\par
\indent

The action for a superparticle system in the pp-wave background 
is obtained by eliminating $\sigma$
dependence in \bref{action}, \bref{C1form}, \bref{dthe}, \bref{psipsi},
\bea
S_{\rm SUPA}&=&\displaystyle\int d\tau
~\frac{1}{2e}{{\bf L}_0}^{\hat{a}}{{\bf L}_0}^{\hat{b}}\eta_{\hat{a}\hat{b}}~~
\eea
where ${{\bf L}_0}^{~\hat{a}}$ is the coefficient of $d\tau$ in the
pullback of the MC form ${\bf L}^{~\hat{a}}$,
\bea
{{\bf L}_0}^{~\hat{a}}&=&
\dot{x}^m {\bf L}_m^{~\hat{a}}+\dot{\theta}^\mu {\bf L}_\mu^{~\hat{a}}~~,~~
{\bf L}_m^{~\hat{a}}\equiv e_m^{~\hat{a}}+\Theta_m^{~~\mu}
{\bf L}_\mu^{~\hat{a}}\label{Lmama}
~~.
\eea
Canonical momenta for $x^m$ and $\theta^\mu$ are
\bea
p_m&\equiv&\displaystyle\frac{\delta S_{\rm SUPA}}{\delta \dot{x}^m}=
\frac{T}{e}{\bf L}_m^{~~\hat{a}}{{\bf L}_0}^{~\hat{b}}\eta_{\hat{a}\hat{b}}
\label{pzeta}\\
\zeta_\mu&\equiv&\displaystyle\frac{\delta^r S_{\rm SUPA}}{\delta \dot{\theta}^\mu}=
\frac{T}{e}{\bf L}_\mu^{~~\hat{a}}{{\bf L}_0}^{~\hat{b}}\eta_{\hat{a}\hat{b}}
\nn~~.
\eea
The gauge invariant Hamiltonian of the pp-superstring is obtained as
\bea
{\cal H}_{\rm SUPA}&=&e H_{\rm SUPA}+F_{\rm SUPA,\mu}\Lambda^\mu
\eea
where primary constraints are
\bea
H_{\rm SUPA}&=&\frac{1}{2}
\pi_{\hat{a}} \pi_{\hat{b}} \eta^{\hat{a}\hat{b}}=0\label{trep}\\
F_{{\rm SUPA},\mu}&=&\zeta_\mu
-\pi_{\hat{a}}{\bf L}_\mu^{~~\hat{a}}=0~~
\label{fsup}
\eea
with super-invariant (up to local Lorentz) combination 
\bea
\pi_{\hat{a}}&=&(e^{-1})_{\hat{a}}^{~~m}
\left(
{p}_m
+{\zeta}_\mu\Theta_{m}^{~~\mu}\right)
=\frac{T}{e}{{\bf L}_0}^{\hat{b}}\eta_{\hat{a}\hat{b}}
\equiv 
(e^{-1})_{\hat{a}}^{~~m}\pi_m.
~~~\label{Ddef}
\eea
Half of the fermionic constraints generates the $\kappa$-symmetry
transformations the existence of which is given 
in the subsection {\bf 3.2}.
Other half is the set of second class constraints and the multipliers 
,$\Lambda^\mu$'s in \bref{fsup},
associating with the second class constraints vanish by the consistency
condition.

Equations of motion in $e=1$ and $\Lambda=0$ gauge 
are obtained in a background covariant way as
\bea
&&\left\{\begin{array}{ccl}
\dot{x}^m&=&\pi^m\\
\dot{p}_m&=&-\frac{1}{2}\partial_m g^{nl}
\pi_n\pi_l
-\zeta (\partial_m\Xi_n) \theta \pi^n
\end{array}\right.\\
&&\left\{\begin{array}{ccl}
\dot{\theta}^\mu&=&\pi^{\hat{a}}(\Xi_{\hat{a}}\theta)^\mu\\
\dot{\zeta}_\mu&=&-\pi^{\hat{a}}(\zeta \Xi_{\hat{a}})_\mu
\end{array}\right.
\eea
with $\Theta_m^{~~\mu}= (\Xi_m)^\mu_{~~\nu}\theta^\nu$
and second order equation is given by 
\bea
\ddot{x}^m=-{\bf \Gamma}^m_{~nl}\dot{x}^n\dot{x}^l
-g^{mn} \dot{x}^l
\zeta (\partial_{[n}\Xi_{l]} -[\Xi_n,\Xi_l] )\theta ~~~.
\eea
where indices $\hat{a}$ and $m$ are raised and lowered by $\eta^{\hat{a}\hat{b}}$ 
and $g^{mn}$ respectively.

The world line element is calculated as
$
ds^2_{\rm SUPA}
=eH_{\rm SUPA}=0
$
where \bref{Ddef} is used.
The superparticle, which is the zero mode of the superstring,
in the pp-wave background moves along the null geodesics
and it satisfies the ``massless" constraint $H_{\rm SUPA}=0$.
.

In order to solve equations of motion the concrete 
expression of the RR pp-wave background metric
\bref{metricg} and 
\bea
\Xi_m=\left\{\begin{array}{l}
\Xi_+=0\\
\Xi_-=\lcpm_+(
2\mu\epsilon\Pi-2\mu^2 x^{\hat{i}}\Gamma_{\hat{i}}\Gamma_+)\\
\Xi_{\hat{i}}=-\mu\epsilon\Pi\Gamma_{\hat{i}}\Gamma_+
\end{array}\right.~~\label{xixi}
\eea
are inserted.
The Hamiltonian is given by
\bea
{\cal H}_{\rm SUPA}&=&\frac{1}{2}
\left[
2p_+(p_-+\zeta \Xi_- \theta) 
+4\mu^2 p_+ {\bmx}^2+(p_{\hat{i}}+\zeta_+ \Xi_{\hat{i}}\theta)^2
\right]\nn\\
&=&
\frac{1}{2}
\left[
2p_+\pi_- +\omega^2 {\bmx}^2+{\bmpi}^2
\right]~~,\label{hamSUPAzeta}
\eea
with $p_m=(p_+,p_-,{\bmp})$, $\pi_m=(\pi_+,\pi_-,{\bmpi})$ 
and $x^m=(x^+,x^-,{\bmx})$. 
The equations of motion become
\bea
&&\left\{\begin{array}{l}
\dot{x}^+=\pi_-+(2\mu)^2 {\bmx}^2p_+\\
\dot{x}^-=p_+\\
\dot{x}^{\hat{i}}={\pi}_{\hat{i}}
\end{array}\right.~,~
\left\{\begin{array}{l}
\dot{p}_-=0\\
\dot{p}_+=0\\
\dot{p}_{\hat{i}}=-\omega^2{x}_{\hat{i}}
+2\mu^2 p_+ \zeta\Gamma_{\hat{i}}\Gamma_+
\end{array}\right.
~\nn\\
&&
\left\{\begin{array}{l}
\dot{\theta}^\mu=p_+(\Xi_-\theta)^\mu+\pi_{\hat{i}}(\Xi_{\hat{i}}\theta)^\mu\\
\dot{\zeta}_\mu=-p_+(\zeta\Xi_-)_{\mu}-\pi_{\hat{i}}(\zeta \Xi_{\hat{i}})_\mu
~~~~~~~~~~~~.
\end{array}\right.\label{eqSUPA}
\eea
Using facts  $\dot{\pi}_-=0,~~\dot{\bmpi}=-\omega^2 {\bmx }$,
the second order equations for $x^m$ take same form as the
bosonic particle \bref{harm}. Furthermore $\theta^\mu$ and $\zeta_\mu$ 
satisfy second order harmonic equation with frequency $\omega$. It is
shown by using
~$\partial_{\hat{i}}\Xi_-=-\Xi_-\Xi_{\hat{i}}$,
~$(\Xi_-)^2=-4\mu^2+4\mu^2 x^{\hat{i}}\Xi_{\hat{i}}$ and 
$\Xi_{\hat{i}}\Xi_{\hat{j}}=0$.
Solutions are found in the same form as \bref{solPA} and
 \bref{hamPA} with replacing
$p_-,p_{\hat{i}}$  by $\pi_-,\pi_{\hat{i}}$. 
The Hamiltonian is expressed as
\bea
{\cal H}_{\rm SUPA}&=&E_B+E_F\label{hambf}\\
E_B&=&\frac{1}{2}
\left[
2p_+\pi_- +\omega^2({\bma}^2+\tilde{\bma}^2)
\right]={\rm const.}~\nn\\
E_F&=&2p_+^2\mu i\bar{\theta}^+\Gamma^+\left(
\displaystyle\frac{\sin \frac{\Psi_+}{2}}{\Psi_+/2}
\right)^2 \epsilon \Pi \theta^+
-2p_+^2\mu^2 i\bar{\theta}^- {\slbmpi}
\left(
\displaystyle\frac{\sin \frac{\Psi_+}{2}}{\Psi_+/2}
\right)^2 {\slbmx}\Gamma^- \theta^-=-{\rm const.}\nn
\eea
where ``$+$" projected part of fermionic constraints \bref{fsup} are used.

In terms of global charges the Hamiltonian can be written as
the quadratic Casimir operator of the super-pp-algebra
which is obtained by the Penrose limit from the
quadratic Casimir operator of the super-AdS algebra;
\bea
&&\begin{array}{l}
c_{\rm sAdS}=\displaystyle\sum
P_{\hat{a}} P_{\hat{b}}\eta^{\hat{a}\hat{b}}+
\displaystyle\sum M_{\hat{a}\hat{b}}^2 
-\frac{1}{4}Q_{\alpha\alpha'A}(\epsilon^{AB}C^{-1\alpha\beta}C'^{-1\alpha'\beta'})
Q_{\beta\beta'B}\\
~~~~~~~~~~~^{\hat{a}=0,1,\cdot,9} ~~~~~~~
^{\hat{a}\hat{b}=0,\cdot,4~{\rm or}~5,\cdots,9} 
\end{array}\nn\\
&&\stackrel{\rm Penrose~limit}{\longrightarrow}~
\begin{array}[t]{l}
c_{{\rm spp}(-2)}=
\displaystyle\sum 
P_{\hat{a}}P_{\hat{b}}\eta^{\hat{a}\hat{b}}
+\displaystyle\sum  {P_{\hat{i}}^*}
{P_{\hat{j}}^*}\eta_*^{\hat{i}\hat{j}}
-\frac{i\eta_*}{4}Q_+({\cal C}^{-1}\Gamma_+\Pi\epsilon )Q_+
~\Rightarrow 2{\cal H}_{\rm SUPA}~~
\\
~~~~~~~~~~~~~^{\hat{a}=0,1,\cdot,9} ~~~~~~~~
^{\hat{i}=1,\cdots,8}
\end{array}\nn
\\~~\label{cas}
\eea
with $\eta_*^{\hat{i}\hat{j}}=(2\sqrt{2} \mu)^2\delta^
{\hat{i}\hat{j}}$ and $\eta_*=2\sqrt{2} \mu$.
The supercharge is obtained by the Penrose limit from the one of the
super-AdS result \cite{AdSWZ}
\bea
Q_+&=&\left\{\zeta_+(1-\frac{i}{2\sqrt{2}}\Gamma_{+\hat{i}}\theta^-
\bar{\theta}^-\Gamma_{\hat{i}}\Pi\epsilon)
-(p_+ i\bar{\theta}^+\Gamma^+ 
+p_{\hat{i}}i\bar{\theta}^-\Gamma^{\hat{i}})
\left(
\displaystyle\frac{\sin \frac{\Psi_+}{2}}{\Psi_+/2}
\right)^2\right\}
\left(
\displaystyle\frac{\Psi_+}{\sin {\Psi_+}}
\right)\nn\\
&&~~~~~~~~~~~~~\times\left(\cos (2\mu x^-) -\epsilon\Pi \sin  (2\mu x^-) 
\right)
\eea
which is consistent with that the Hamiltonian \bref{hamSUPAzeta}
and the quadratic Casimir operator in \bref{cas}.

\subsection{ Bosonic string}\par
\indent

The action for a bosonic string system is 
obtained by eliminating $\theta$ dependence in \bref{Action},
\bea
S_{\rm ST}&=&-~T\displaystyle\int d^2\sigma~
\sqrt{-h}h^{\uu\vv}~\partial_\uu x^m g_{mn}
\partial_\vv x^n\nn
~~.
\eea
Canonical momentum of $x^m(\sigma)$ is 
$p_m(\sigma)=\delta S_{\rm ST}/\delta \dot{x}^m(\sigma)=
-T\sqrt{-h}h^{u0}\partial_u x^n g_{mn}$.
The gauge invariant Hamiltonian of the pp-string is obtained as
\bea
{\cal H}_{\rm ST}&=&\displaystyle\int d\sigma~
\left(\frac{\sqrt{-h}}{h_{11}}{ H}_{\perp}+
\frac{h_{01}}{h_{11}}{ H}_\parallel\right),\\
\eea
where primary constraints are
\bea
H_\perp(\sigma)&=&\frac{1}{2T}(p_mg^{mn}p_n
+T^2 {x'}^mg_{mn}{x'}^n)
=0\\
H_{{\parallel}}(\sigma)&=&p_m {x'}^m=0~~.
\eea
The equation of motion in the conformal gauge, 
$h_{\uu\vv}=\eta_{\uu\vv}$,
are obtained as
\bea
\left\{\begin{array}{ccl}
\dot{x}^m(\sigma)&=&\frac{1}{T}g^{mn}p_n\\
\dot{p}_m(\sigma)&=&T(g_{mn}{x'}^{n})'
+\frac{T}{2}(\partial_mg_{nl})\left(\frac{1}{T^2}p^np^l-{x'}^n{x'}^l\right)
\end{array}\right.
\eea
and
\bea
\Box{x}^m(\sigma)
=-{\bf \Gamma}^m_{~nl}(\dot{x}^n\dot{x}^l-{x'}^n{x'}^l)~~
\eea
with $\Box=\partial_\tau^2-\partial_\sigma^2=\partial_0^2-\partial_1^2$.

For the RR pp-wave background case, by using
\bref{metricg}, the Hamiltonian becomes
\bea
{\cal H}_{\rm ST}&=&
\frac{1}{2T}\displaystyle\int d\sigma~
\left[
2p_+p_- +{\bmp}^2
+(2\mu p_+) ^2{\bmx}^2
+T^2\left(
{x^+}'{x^-}'+{{\bmx}'}^2
-(2\mu)^2{\bmx}^2  {{x^-}'}^2\right) 
\right]~.\nn\\
\label{hamST}~~
\eea
The equations of motion are
\bea
\left\{\begin{array}{l}
\dot{x}^+(\sigma)=\frac{1}{T}(p_-+(2\mu)^2 {\bmx}^2p_+)\\
\dot{x}^-(\sigma)=\frac{1}{T}p_+\\
\dot{\bmx}(\sigma)=\frac{1}{T}{\bmp}
\end{array}\right.~~,~~
\left\{\begin{array}{l}
\dot{p}_-(\sigma)=T({x^+}''-(2\mu)^2 ({\bmx}^2 {x^-}')')\\
\dot{p}_+(\sigma)=T{x^-}''\\
\dot{\bmp}(\sigma)=-T\left\{{\bmx}''+
+(2\mu )^2{\bmx}\left((\frac{p_+}{T})^2-({x^-}')^2
\right)  \right\}
\end{array}\right.\label{eqST}~~,
\eea
and ones of the second order form become
\bea
\left\{\begin{array}{l}
\Box x^+(\sigma)=(2\mu)^2 \dot{{\bmx}^2}\dot{x}^-\\
\Box x^-(\sigma)=0\\
\Box {\bmx}(\sigma)=(2\mu )^2 {\bmx}\left(({x^-}')^2-(\dot{x}^-)^2\right)
~~.\end{array}\right.~~\label{eqSTST}~~
\eea 
By solving equation of motion for $x^-$ in \bref{eqSTST} as
\bea
x^-(\sigma)=\frac{p_{+,0}}{T}\tau+
\frac{1}{\sqrt{2\pi}}\displaystyle\sum_{n \neq 0}
\left(\alpha^-_n e^{2in(\sigma-\tau)}
+\tilde{\alpha}^-_n e^{-2in(\sigma+\tau)}\right)\label{xminus}~~,
\eea
the last term of the last equation in \bref{eqSTST} becomes
\bea
({x^-}')^2-(\dot{x}^-)^2
=-\left(\frac{p_{+,0}}{T}-4i\displaystyle\sum_{n \neq 0}
n\alpha^-_n e^{2in(\sigma-\tau)}\right)
\left(\frac{p_{+,0}}{T}-4i\displaystyle\sum_{n \neq 0}
n\tilde{\alpha}^-_n e^{-2in(\sigma+\tau)}
\right)~~~.
\eea
It is difficult to solve the equation for ${\bmx}$ 
except in a case where only the zero-mode term 
 $(p_{+,0}/T)^2$ is present. 
It is the same as the light-cone result \cite{MeTsy}
\bea
&&{\bmx}={\bma}_0 \sin (\omega_0 \tau)
+\tilde{\bma}_0 \cos (\omega_0 \tau)
+\frac{1}{\sqrt{2\pi}}\displaystyle\sum_{n \neq 0}
\left({\bma}_n e^{i(\omega_n\tau +2n\sigma)}
+\tilde{\bma}_n e^{i(\omega_n\tau-2n\sigma)}\right)
~~\nn\\
&&~~~~~~~~~~~~~~~~~\omega_n={\rm sgn}(n)\sqrt{(2\mu p_{+,0})^2 +(2n)^2}\label{xtran}
\eea  
with ${\bma}_n^\dagger={\bma}_{-n}$ and  $\tilde{\bma}_n^\dagger=\tilde{\bma}_{-n}$
and the Hamiltonian becomes 
\bea
{\cal H}_{\rm ST}&=&\frac{1}{2T}
\left[
\omega_0^2 ({\bma}_0^2+\tilde{\bma}_0^2)
+\displaystyle\sum_{n \neq 0}
\omega_n^2 {\bma}_n{\bma}_{-n}
+\cdots
\right]\label{hamSTSTST}~~.
\eea

\subsection{ Superstring}\par
\indent

Now let us examine a superstring in the RR pp-wave background
which is described by the covariant action \bref{Action}.
The conjugate momenta are
\bea
p_m(\sigma)&\equiv &\displaystyle\frac{\delta {\cal L}}{\delta \dot{x}^m}
=\displaystyle\frac{\delta {\cal L}_0}{\delta {\bf L}_0^{~~\hat{a}}}
{\bf L}_m^{~~\hat{a}}
+\displaystyle\frac{\delta {\cal L}_{WZ}}{\delta L_0^{~~\alpha}}L_m^{~\alpha}
\\
\zeta_\mu(\sigma)&\equiv&\displaystyle\frac{\delta^r {\cal L}}{\delta \dot{\theta}^\mu}
=\displaystyle\frac{\delta {\cal L}_0}{\delta {\bf L}_0^{~~\hat{a}}}
{\bf L}_\mu^{~~\hat{a}}
+\displaystyle\frac{\delta^r {\cal L}_{WZ}}{\delta L_0^{~~\alpha}}L_\mu^{~\alpha}
~~.
\eea
The gauge invariant Hamiltonian of the pp-superstring is obtained as
\bea
{\cal H}&=&
\frac{\sqrt{-h}}{h_{11}}{\cal H}_{\perp}+\frac{h_{01}}{h_{11}}{\cal H}_{\parallel}
\nn\\
{\cal H}_{\perp}(\sigma)&=&H_{\perp}-F\tau_3 \theta'\\
{\cal H}_{\parallel}(\sigma)&=&H_{\parallel}+F\theta'\nn
\eea
where primary constraints are
\bea
H_\perp (\sigma)&=&\frac{1}{2T}(
\tilde{\pi}_{\hat{a}} \tilde{\pi}_{\hat{b}} \eta^{\hat{a}\hat{b}}
+T^2{\bf L}_1^{~\hat{a}}{\bf L}_1^{~\hat{b}}\eta_{\hat{a}\hat{b}})=0
\label{Hperp}\\
H_{{\parallel}} (\sigma)&=&\tilde{\pi}_{\hat{a}}{\bf L}_1^{~\hat{a}}=0
\label{Hparallel}\\
F_{\nu} (\sigma)&=&\zeta_\nu
-\tilde{\pi}_{\hat{a}}{\bf L}_\nu^{~~\hat{a}}
-\displaystyle\frac{\delta {\cal L}_{WZ}}{\delta L_0^\beta}L_\nu^{~~\beta}=0
\eea
Half of the above fermionic constraints generate
$\kappa$-symmetry as shown in the section 3.2 
for the action \bref{Action}.

Super-invariant (up to local Lorentz) combinations are ${\bf L}_1^{\hat{a}}$ and
\bea
\tilde{\pi}_{\hat{a}}&=&(e^{-1})_{\hat{a}}^{~~m}
(\tilde{p}_m
+\tilde{\zeta}_\mu\Xi_m\theta^{\mu})
=-\frac{T}{\sqrt{-G}}(-{{\bf L}_0}^{\hat{b}} G_{11}+{\bf L}_1^{\hat{b}} G_{01})\eta_{\hat{b}\hat{a}}
\equiv (e^{-1})_{\hat{a}}^{~~m}
\tilde{\pi}_m
\nn\\
&&~~~\tilde{p}_m=p_{m}-
\displaystyle\frac{\delta {\cal L}_{WZ}}{\delta L_0^\beta}
L_{m}^{~~\beta}~~,~~
\tilde{\zeta}_\mu=\zeta_\mu-
\displaystyle\frac{\delta {\cal L}_{WZ}}{\delta L_0^\beta}
L_{\mu}^{~~\beta}\nn~~.
\eea

The equations of motion for a superstring in the pp-wave background 
in the conformal gauge, 
$h_{\uu\vv}=\eta_{\uu\vv}$,
are obtained as
\bea
&&\left\{\begin{array}{ccl}
\dot{x}^m&=&\frac{1}{T}\tilde{\pi}^m
+(e^{-1})_{\hat{a}}^{~~m}
{\bf L}_\mu^{~~\hat{a}}(\tau_3 \theta')^\mu
\\
\dot{p}_m&=&T\left({\bf L}_m^{~~\hat{a}}
{\bf L}_{1\hat{a}}\right)'\\
&&
+(\partial_mg_{nl})
\left(\frac{1}{2T}\tilde{\pi}^n\tilde{\pi}^l
-\frac{T}{2}{\bf L}_1^{~~\hat{a}}(e^{-1})_{\hat{a}}^{~~n}
{\bf L}_1^{~~\hat{b}}(e^{-1})_{\hat{b}}^{~~l}
-\tilde{\pi}^n{\bf L}_\mu^{~~\hat{a}}(e^{-1})_{\hat{a}}^{~~l}
(\tau_3 \theta')^\mu
\right)\\
&&
-\frac{1}{T}{\zeta}(\partial_m\Xi_n)\theta \tilde{\pi}^n
-T {x^n}'\left((\partial_m\Xi_n)\theta\right)^\mu 
{\bf L}_\mu^{~~\hat{a}}{\bf L}_{1\hat{a}}\\
&&
-\partial_m
\left(
 \displaystyle\frac{\delta {\cal L}_{WZ}}{\delta L_0^{~~\alpha}}\right)L_\mu^{~~\alpha}
( \tau_3\theta')^\mu
\end{array}\right.\nn\\\\
&&\left\{\begin{array}{ccl}
\dot{\theta}^\mu&=&
-(\tau_3 \theta')^\mu
+(\Xi_{m}\theta)^\mu
\left(\frac{1}{T}\tilde{\pi}^m+(e^{-1})_{\hat{a}}^{~~m}{\bf L}_\nu^{~~\hat{a}}
(\tau_3 \theta')^\nu
\right)
\\
\dot{\zeta}_\mu&=&-(\zeta \tau_3)'_\mu
-T({\bf L}_\mu^{~~\hat{a}}{\bf L}_{1\hat{a}})'
-(\tilde{\pi}_{\hat{a}}{\bf L}_\mu^{~~\hat{a}}\tau_3)'
\\&&
-(\zeta\Xi_{\hat{a}})_{\mu}
\left(\frac{1}{T}\tilde{\pi}^{\hat{a}}+{\bf L}_\nu^{~~\hat{a}}(\tau_3 \theta')^\nu
\right)\\
&&+T\left({x'}^m(\Xi_m)^\nu_{~~\mu}{\bf L}_\nu^{~~\hat{a}}
-({x'}^m (\Xi_m\theta)^\nu+{\theta'}^\nu)
\displaystyle\frac{\partial^{\rm left}}{\partial \theta^\mu}{\bf L}_\nu^{~~\hat{a}}
\right){\bf L}_{1\hat{a}}\\
&&+\displaystyle\frac{\partial^{\rm left}}{\partial \theta^\mu}\left(
\displaystyle\frac{\delta {\cal L}_{WZ}}{\delta L_0^{~~\alpha}}L_\nu^{~~\alpha}\right)
(\tau_3 \theta')^\nu
+\tilde{\pi}_{\hat{a}}\left(\displaystyle\frac{\partial^{\rm left}}{\partial \theta^\mu} {\bf L}_\nu^{~~\hat{a}}\right)(\tau_3\theta')^\nu
\end{array}\right.
\eea
which are background covariant.

In the components by using \bref{metricg} and \bref{xixi}
the Hamiltonian in the conformal gauge is rewritten as
\bea
{\cal H}&=&{\cal H}_1+{\cal H}_2+{\cal H}_3\label{hmsustcp}\\
{\cal H}_{1}&=&\displaystyle\int d\sigma~\left[
\displaystyle\frac{1}{2T}\left(
2p_+ \tilde{\pi}_- +4\mu^2{\bmx}^2p_+^2 +\tilde{\bmpi}^2
\right)\right.\nn\\
&&~~~~\left.+p_+{\bf L}_\mu^{~~+}(\tau_3 \theta')^\mu
+(\tilde{\pi}_- +2\mu^2{\bmx}^2p_+){\bf L}_{\nu}^{~~-}
(\tau_3{\theta^-}')^{\nu}
+\tilde{\pi}_{\hat{i}}{\bf L}_\mu^{~~\hat{i}}(\tau_3\theta')^\mu
\right]\nn\\
{\cal H}_{2}&=&\displaystyle\int d\sigma~\left[
\displaystyle\frac{T}{2}\left\{
2\left(
({x^+}'-2\mu^2{\bmx}^2{x^-}')+(\theta'+{x^-}'\Xi_-\theta
+{x^{\hat{i}}}'\Xi_{\hat{i}}\theta^-)^\mu
{\bf L}_\mu^{~~+}
\right)
({x^-}'+({\theta^-}')^{\nu'}{\bf L}_{\nu}^{~~-})
\right.\right.\nn\\&&~~~~
\left.\left.+\left(
{x^{\hat{i}}}'+(\theta'+{x^-}'\Xi_-\theta 
+{x^{\hat{i}}}'\Xi_{\hat{i}}\theta^-)^\mu{\bf L}_\mu^{~~\hat{i}}
\right)^2 \right\} +\displaystyle\frac{\delta {\cal L}_{WZ}}{\delta L_0^
{~~\alpha}}
L_\mu^{~~\alpha}(\tau_3 \theta')^\mu    \right]\nn\\
{\cal H}_{3}&=&\displaystyle\int d\sigma~\left[
-\zeta \tau_3 \theta'
\right]\nn~~.
\eea
For a flat case ${\cal H}_1$ and ${\cal H}_2$ 
reduce into simply $\displaystyle\int (1/2T)p^2$ and $\displaystyle\int (T/2) {x'}^2$
respectively. 
Equations of motion are written as
\bea
&&\left\{\begin{array}{ccl}
\dot{x}^+&=&\frac{1}{T}(\tilde{\pi}_-+4\mu^2{\bmx}^2p_+)
+{\bf L}_\mu^{~~+}(\tau_3\theta')^\mu
+2\mu^2 {\bmx}^2{\bf L}_\nu^{~~-}(\tau_3{\theta^-}')^\nu\\
\dot{x}^-&=&\frac{1}{T}p_++{\bf L}_\nu^{~~-}(\tau_3 {\theta^-}')^\nu\\
\dot{x}^{\hat{i}}&=&\frac{1}{T}\tilde{\pi}_{\hat{i}}+{\bf L}_\mu^{~~\hat{i}}
(\tau_3\theta')^\mu
\end{array}\right.\\
&&\left\{\begin{array}{ccl}
\dot{p}_-&=&T\left\{
-2\mu^2{\bmx}^2{\bf L}_1^{~~-}
+(\Xi_-\theta)^\mu ({\bf L}_\mu^{~~+}{\bf L}_1^{~~-}
+{\bf L}_\mu^{~~\hat{i}}{\bf L}_1^{~~\hat{i}})
\right\}'\\
&&+\int d\sigma'
\frac{
\partial}{\partial x^-(\sigma)}\left(
\frac{\delta {\cal L}_{WZ}}{\delta L_0^{~~\alpha}}\right)
\left\{2L_\mu^{~~\alpha}
\left(\lcpm_+(\Xi_-\theta \dot{x}^-+
\Xi_{\hat{i}}\theta\dot{x}^{\hat{i}})\right)^\mu
-L_\mu^{~~\alpha}(\tau_3\theta')^\mu
\right\}\\
\dot{p}_+&=&T\left({x^-}''+(({\theta^-}')^\nu{\bf L}_\nu^{~~-})'
\right)+
iT\left(
(\frac{\sin \Psi_-}{\Psi_-}\epsilon\Pi\theta^-)^T{\cal C}\Gamma_+\tau_1\Pi
\frac{\sin \Psi_-}{\Psi_-}\tau_3{\theta^-}'
\right)'\\
\dot{p}_{\hat{i}}&=&+T\left({\bf L}_1^{~~\hat{i}}
+(\Xi_{\hat{i}}\theta^-)^\mu({\bf L}_\mu^{~~+}{\bf L}_1^{~~-}
+{\bf L}_\mu^{~~\hat{i}}{\bf L}_1^{~~\hat{i}})
\right)'\\
&&-4\mu^2 x^{\hat{i}}\left\{ p_+\left(\frac{1}{T}p_+
+{\bf L}_\nu^{~~-}(\tau_3{\theta^-}')^\nu\right)-
{x^-}'{\bf L}_1^{-}\right\}\\
&&-T{x^-}'(\Xi_-\Xi_i\theta^-)^\mu({\bf L}_\mu^{~~+}{\bf L}_1^{~~-}
+{\bf L}_\mu^{~~\hat{i}}{\bf L}_1^{~~\hat{i}}
)-
\int d\sigma'
\left(
\frac{\partial}{\partial x^{\hat{i}}(\sigma)}
\frac{\delta {\cal L}_{WZ}}{\delta L_0^{~~\alpha}}
\right)L_\mu^{~~\alpha}(\tau_3 \theta')^\mu\\
&&-\frac{iT}{\mu}\left\{
\left(\frac{\sin\Psi_+}{\Psi_+}\Xi_{\hat{i}}\theta^-\right)^T
{\cal C}\Gamma_-\tau_1\Pi\frac{\sin \Psi_+}{\Psi_+}
\left(\Xi_-\theta\dot{x}^- +\Xi_{\hat{j}}\theta^-\dot{x}^{\hat{j}}
\right)
\right\}'\\
&&-\frac{iT}{\mu}\left\{
\left(\frac{\sin\Psi_+}{\Psi_+}{x^-}'\Xi_-\Xi_{\hat{i}}\theta^-\right)^T
{\cal C}\Gamma_-\tau_1\Pi\frac{\sin \Psi_+}{\Psi_+}
\left(\Xi_-\theta\dot{x}^- +\Xi_{\hat{j}}\theta^-\dot{x}^{\hat{j}}
\right)
\right\}
\end{array}\right.\\
&&\left\{\begin{array}{ccl}
(\dot{\theta}^+)^\lambda&=&-(\tau_3{\theta^+}')^\lambda
+(\Xi_-\theta)^\lambda\left(
\frac{1}{T}p_+
+{\bf L}_\nu^{~~-}(\tau_3 {\theta^-}')^\nu\right)
+(\Xi_{\hat{i}}\theta^-)^\lambda
\left(\frac{1}{T}\tilde{\pi}_{\hat{i}}
+{\bf L}_\mu^{~~\hat{i}}
(\tau_3\theta')^\mu\right)\\
\dot{\theta}^-&=&-\tau_3{\theta^-}'\\
(\dot{\zeta}_+)_\lambda&=&-({\zeta_+}'\tau_3)_\lambda
-(\zeta_+-\frac{iT}{\mu}\bar{L}_1^{~+}\Gamma_-\tau_1\Pi\frac{\sin \Psi_+}{\Psi_+})_\nu
(\Xi_-\lcpm_+)^\nu_{~~\lambda}
\left(\frac{1}{T}p_++{\bf L}_\mu^{~~-}(\tau_3 {\theta^-}')^\mu\right)\\
&&+T\left[-({\bf L}_\mu^{~~+}{\bf L}_1^{~~-}+{\bf L}_\mu^{~~\hat{i}}{\bf L}_1^{~~\hat{i}})'(\lcpm_+)^\mu_{~~\lambda}
+({x^-}'\Xi_-)^\mu_{~~+}({\bf L}_\mu^{~~+}{\bf L}_1^{~~-}
+{\bf L}_\mu^{~~\hat{i}}{\bf L}_1^{~~\hat{i}})\right.\\
&&\left.
-(\theta'+{x^-}'\Xi_-\theta+{x^{\hat{i}}}'\Xi_{\hat{i}}\theta^-)^\mu
\left\{\left(
\frac{\partial^{\rm left}}{\partial (\theta^+)^\lambda}{\bf L}_\nu^{~~+}
\right){\bf L}_1^{~~-}
+\left(
\frac{\partial^{\rm left}}{\partial (\theta^+)^\lambda}{\bf L}_\nu^{~~\hat{j}}
\right){\bf L}_1^{~~{\hat{j}}}
\right\}\right]\\
&&+(p_+{\bf L}_\nu^{~~+}(\lcpm_+)^\nu_{~\lambda}\tau_3 +\tilde{\pi}_{\hat{i}}{\bf L}_\nu^{~~\hat{i}}(\lcpm_+)^\nu_{~~\lambda}\tau_3)'\\&&
+\left(
\frac{\partial^{\rm left}}{\partial (\theta^+)^\lambda}{\bf L}_\nu^{~~+}\right)
(\tau_3\theta')^\nu
p_+
+\left(
\frac{\partial^{\rm left}}{\partial (\theta^+)^\lambda}{\bf L}_\nu^{~~\hat{j}}\right)
(\tau_3\theta')^\nu
\tilde{\pi}_{\hat{i}}\\&&
+\frac{iT}{2\mu}\int d\sigma'\left(
\frac{\partial^{\rm left}}{\partial (\theta^+)^\lambda(\sigma)}
(\bar{L}_1^{~+}\Gamma_-\tau_1\Pi)_+L_\mu^{~~+}(\tau_3\theta')^\mu\right)(\sigma')\\&&
+\frac{iT}{\mu}\left\{
\left(\frac{\sin\Psi_+}{\Psi_+}\right)^T
{\cal C}\Gamma_-\tau_1\Pi\frac{\sin \Psi_+}{\Psi_+}
\left(\Xi_-\theta\dot{x}^- +\Xi_{\hat{j}}\theta^-\dot{x}^{\hat{j}}
\right)
\right\}'\\
&&-\frac{iT}{\mu}\left\{
\left(\frac{\sin\Psi_+}{\Psi_+}{x^-}'\Xi_-\right)^T
{\cal C}\Gamma_-\tau_1\Pi\frac{\sin \Psi_+}{\Psi_+}
\left(\Xi_-\theta\dot{x}^- +\Xi_{\hat{j}}\theta^-\dot{x}^{\hat{j}}
\right)
\right\}
\end{array}\right.
\eea
and second order form equations become
\bea
\left\{\begin{array}{l}
\Box x^+(\sigma)=(2\mu)^2 \dot{{\bmx}^2}\dot{x}^-+\theta^\pm~{\rm dependent~terms}\\
\Box x^-(\sigma)=
\partial_\tau\left({\bf L}_\nu^{~~-}(\tau_3{\theta^-}')^\nu\right)
-\partial_\sigma\left({\bf L}_\nu^{~~-}({\theta^-}')^\nu\right)
+i\left\{
(\frac{\sin\Psi_-}{\Psi_-}\epsilon\Pi\theta^-)^T{\cal C}\Gamma_+\tau_1\Pi
\frac{\sin\Psi_-}{\Psi_-}\tau_3{\theta^-}'
\right\}'
\\
\Box {\bmx}(\sigma)=(2\mu )^2 {\bmx}\left(({x^-}')^2-(\dot{x}^-)^2\right)
+\theta^-~{\rm dependent~terms}
\end{array}\right.~~\label{eqSUST}~~.
\eea 
The right hand sides of the first and third 
equations are complicated functions of $\theta^-$
or $\theta^-$ and $\theta^+$.

\section{ Conclusions and discussions}\par

In this paper a form of the gauge invariant action for the superstring 
in the RR pp-wave 
background is proposed. It is explicit since the Wess-Zumino term is 
bilinear with respect to the LI currents of the super-pp-wave algebra.
It is obtained by the Penrose limit from the superstring action in the 
AdS background with the bilinear WZ term \cite{AdSWZ}.
The Penrose limit of 
the WZ term is given essentially as follows:
\begin{enumerate}
\item{Rescale the coordinates in the LI 1-forms 
with a parameter $\Omega$ with suitable weights, for example 
\bea
L_{\rm AdS}^-(z^m\to \Omega^{N_m}z^m)~
&=&~L^-_{(0)}(z^m)+\Omega^2 L_{(2)}^-(z^m)+o(\Omega^4)~~.
\eea}
\item{Rescale 
the WZ term with the same weight of the Nambu-Goto term as
\bea
{\cal L}_{WZ,{\rm AdS}}&\to& \Omega^{-2}{\cal L}_{WZ,{\rm AdS}}~~~.
\eea
The term which would have negative power of $ \Omega$ is
$\bar{L}_{\rm AdS}^- \Gamma^-\Pi\tau_1 L_{{\rm AdS}}^-$ 
and becomes
\bea
\Omega^{-2}\bar{L}^-_{\rm AdS} \Gamma^-\Pi\tau_1 L^-_{{\rm AdS}}~&=&~
\Omega^{-2}\left(
\bar{L}^-_{(0)} \Gamma^-\Pi\tau_1 L^-_{(0)}
+\Omega^2~2\bar{L}^-_{(0)} \Gamma^-\Pi\tau_1 L^-_{(2)}
+o(\Omega^4)\right)~~.\nn
\eea}
\item{Subtract the divergent term in $\Omega\to 0$ limit
which is proportional to $1/\Omega^2$ and closed.}
\item{Take the $\Omega\to 0$ limit in the bilinear WZ term
\bea
{\cal L}_{WZ,{\rm AdS}}&\to&
{\cal L}_{WZ,{\rm pp}}=2\bar{L}^-_{(0)} \Gamma^-\Pi\tau_1 L^-_{(2)}+
L^+~{\rm dependent~term}~~.\nn
\eea}
\end{enumerate}
This procedure corresponds to make a nondegenerate   super-pp-wave group by
introducing the fermionic center associating to $L^-_{(2)}=\tilde{L}^-$.
This is contrasted with the conventional WZ term case:
\begin{enumerate}
\item{Rescale coordinates in the LI 1-forms 
with a parameter $\Omega$
with suitable weights and take the leading terms in the limit 
$\Omega \to 0$
\bea
L_{\rm AdS}^-(z^m\to \Omega^{N_m}z^m)~
\to~L^-_{(0)}(z^m)~~,~~{\bf L}_{\rm AdS}^+(z^m\to \Omega^{N_m}z^m)~
\to~\Omega^2 {\bf L}^+_{(2)}(z^m).
\eea}
\item{Construct the  WZ term in  the conventional form 
given by an integral of the three form \cite{Me1}
\bea
\displaystyle\int d^2\sigma
{\cal L}_{WZ,{\rm AdS}}&\to&\displaystyle\int d^2\sigma
 {\cal L}_{WZ,{\rm pp}}=\displaystyle\int d^3\sigma
[\bar{L}_{(0)}^- \slbL_{(2)}^+ \tau_3 L_{(0)}^- +L^\pm~{\rm dependent~terms}]~~.
\nn~~
\eea}
\end{enumerate} 
In this case the next to leading term $L_{(2)}^-$ is not 
necessary to be preserved                       
in taking the  Penrose limit of the WZ term.    
However the resultant WZ term does not allow us simple treatment
of the Hamiltonian and equations of motion.

The Hamiltonians for the bosonic particle, superparticle,
string and superstring in the RR pp-wave background are
obtained in the conformal gauge.
The particle and the superparticle Hamiltonians are
 identified with the quadratic Casimir operators
of the pp-wave and the super-pp-wave algebras respectively. 
Once the superparticle Hamiltonian \bref{trep} is recognized as
the super-pp-invariant ``mass operator" 
of the superstring theory,
all  states in a supergravity multiplet are ``massless" 
and the supersymmetry is manifest.

The world-sheet reparametrization generators and the 
local fermionic constraints are also obtained.
The combinations of the reparametrization constraints
\bref{Hperp} and \bref{Hparallel} as
 ${\cal H}_\perp \pm {\cal H}_\parallel$ 
and the first class part of $F_\nu^I$, where $I=1,2$ correspond to right/left
($\pm$) modes, will make a closed set of constraint algebra, 
namely ${\cal ABCD}$ constraint
system \cite{ABCD}. 
It was shown that the local constraints of the AdS superstring
satisfy the ${\cal ABCD}$ algebra \cite{HKAdS}
as well as of the flat superstring.
Since the super-pp-wave algebra is obtained by the Penrose limit
from the super-AdS algebra \cite{HKSpp}
as well as the flat algebera by the flat limit,
the local symmetry algebra is also expected to be obtained 
by the same limiting procedure
preserving the same structure of the ${\cal ABCD}$ algebra.    
The background independence of the local symmetries
is plausible.

Equations of motion in the conformal gauge are obtained 
and are background covariant.
The equations of motion for $x^\pm$
are obtained and will be important for 
taking into account interactions.
In order to quantize the superstring theory in the conformal gauge 
there may be required a suitable change of variables such as to the 
GL(4$\mid$4) matrix variable as was done for the AdS$_5\times$S$^5$ case
\cite{RWS}. 
The covariant approach will be  useful to examine 
symmetry structures toward the covariant superstring field theory, 
S-T-U dualities 
which are deeply related to the background symmetry
and non-perturbative properties such as BPS conditions.  

\appendix
\section{ Cartan 1-forms for AdS$_5\times$S$^5$}\par

The Cartan 1-forms in the AdS$_5\times$S$^5$ space 
are presented \cite{MeTsy,AdSWZ}.
The left-invariant Cartan one-forms of a coset $
{\rm SU}(2,2|4)/[{\rm SO}(4,1)\times{\rm SO}(5)]~\ni~
G=G(x,\theta)=e^{x P}e^{\theta Q} \nn
$
are defined by $
G^{-1}dG={\bf L}^aP_a+{\bf L}^{a'}P_{a'}+\frac{1}{2}{\bf L}^{ab}J_{ab}+\frac{1}{2}{\bf L}^{a'b'}J_{a'b'}
+L^{\alpha\alpha'I}Q_{\alpha\alpha' I}$. 
They are given by
\bea
\left\{\begin{array}{lcl}
{\bf L}^a=e^a+i\theta CC'\gamma^a~
    \left(\displaystyle\frac{\sin (\frac{\Psi}{2})}{\Psi/2}\right)^2~D\theta~~~~~~&,&
{\bf L}^{a'}=e^{a'}-\theta CC'\gamma^{a'}~
    \left(\displaystyle\frac{\sin (\frac{\Psi}{2})}{\Psi/2}\right)^2~D\theta\\
{\bf L}^{ab}=\omega^{ab}-\theta CC'\gamma^{ab}\epsilon~
    \left(\displaystyle\frac{\sin (\frac{\Psi}{2})}{\Psi/2}\right)^2~D\theta&,&
{\bf L}^{a'b'}=\omega^{a'b'}+\theta CC'\gamma^{a'b'}\epsilon~
    \left(\displaystyle\frac{\sin (\frac{\Psi}{2})}{\Psi/2}\right)^2~D\theta\\
    L^{\alpha}=\displaystyle\frac{\sin \Psi}{\Psi}~D\theta~~,~~~~~~~~~~~~~~~~~~~~~
   &&\\
\omega^{ab}=\frac{1}{2}\left(\displaystyle\frac{\sinh (\frac{x}{2})}{x/2}\right)^2 dx^{[a}x^{b]}
~~~~~~~~~~~~~~~&,&
\omega^{a'b'}=-\displaystyle\frac{1}{2}\left(\displaystyle\frac{\sin (\frac{x'}{2})}{x'/2}\right)^2 dx^{[a'}x^{b']}
\end{array}\right.
\label{LI1}
\eea
where $[ab]=ab-ba$ 
and charge conjugation matrix for AdS$_5$ space and S$^5$ space are $C$ and $C'$ respectively and  
\bea
 D\theta&=&\left[d-\frac{i}{2}\epsilon(\gamma^ae_a+i\gamma^{a'}e_{a'})
    	+\frac{1}{4}(\gamma^{ab}\omega_{ab}+\gamma^{a'b'}\omega_{a'b'})\right]\theta\nn\\ 
 (\Psi^2)^{\alpha\alpha'I}_{~~{\beta\beta' J}}&=&(\epsilon \gamma^{{a}}\theta)^{\alpha\alpha'I}~
(\theta CC'\gamma_{{a}})_{\beta\beta' J}-(\epsilon\gamma^{{a'}}\theta)^{\alpha\alpha'I}~
(\theta CC'\gamma_{{a'}})_{\beta\beta' J}\\
&&-\frac{1}{2}(\gamma^{{a}{b}}\theta)^{\alpha\alpha'I}~
(\theta CC'\gamma_{{a}{b}}\epsilon)_{\beta\beta' J}
+\frac{1}{2}(\gamma^{{a'}{b'}}\theta)^{\alpha\alpha'I}~
(\theta CC'\gamma_{{a'}{b'}}\epsilon)_{\beta\beta' J}\nn
~~.
\eea
After the Penrose limit using \bref{Penvar} and 
\bref{PLtheta} 
they are written as \bref{C1form}.
The relation between the AdS variables and
 the Penrose variables are described in \cite{HKSpp}.


\end{document}